\documentclass{WileyMSP-template}

\usepackage{amsmath,bm}
\usepackage{amssymb}
\usepackage{amsfonts}
\usepackage{graphicx}
\usepackage[T1]{fontenc}
\usepackage{textcomp}

\usepackage[sort, nocompress]{cite}
\usepackage[version=4]{mhchem}

\usepackage{color}
\usepackage{braket}
\usepackage{physics}
\newcommand{\braind}[2]{{}^{}_{#1}\!\bra{#2}}
\begin{document}

	\pagestyle{fancy}
	\rhead{\includegraphics[width=2.5cm]{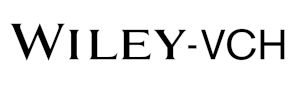}}

	\title{How to read out the phonon number statistics via resonance fluorescence spectroscopy of a single-photon emitter}
	
	\maketitle

	
	\author{Daniel Groll*}
	\author{Fabian Paschen}
	\author{Pawe\l{} Machnikowski}
	\author{Ortwin Hess}
	\author{Daniel Wigger}
	\author{Tilmann Kuhn*}

	\begin{affiliations}
		Daniel Groll, Fabian Paschen, Prof. Tilmann Kuhn\\
		Institute of Solid State Theory, University of M\"unster, 48149 M\"unster, Germany\\
		
		Prof. Pawe\l{} Machnikowski\\
		Institute of Theoretical Physics, Wroc\l{}aw University of Science and Technology, 50-370 Wroc\l{}aw, Poland\\
		
		Prof. Ortwin Hess, Dr. Daniel Wigger\\
		School of Physics, Trinity College Dublin, Dublin 2, Ireland\\
		
		Prof. Ortwin Hess\\
		CRANN Institute and Advanced Materials and Bioengineering Research (AMBER), Trinity College Dublin, Dublin 2, Ireland\\
		
		Email Addresses: daniel.groll@uni-muenster.de, tilmann.kuhn@uni-muenster.de	
	\end{affiliations}

	

	\begin{abstract}
		
		In today's development of quantum technologies a hybrid integration of phononic excitations becomes increasingly attractive. As natural quasi-particle excitations in solid state systems, phonons couple to virtually any other excitation and therefore constitute a useful interaction channel between different building blocks in hybrid quantum systems. This work explores how the efficient light-scattering properties of a single-photon emitter and the appearance of characteristic sidebands in resonance fluorescence spectra, when interfaced with an arbitrary phonon quantum state, can be utilized for acousto-optical transduction. Within reasonable approximations, an analytical description for the optical spectra in the low excitation limit is developed which can be used to read the number statistics of the initial phonon state from a given spectrum. It is shown that the readout is faulty in situations where relevant resonant transitions are forbidden due to vanishing Franck-Condon factors, especially when considering spectra with a noisy background. Two possible solutions to this problem are presented: (A) changing the detuning of the laser relative to the single-photon emitter which modifies the relevant resonant transitions, or (B) increasing dissipation of the single-photon emitter to promote off-resonant transitions. 
		
	\end{abstract}
	
	
	\section{Introduction}
	Current approaches to quantum technology utilize single photon emitters (SPEs) in the form of quantum dots or color centers as basic building blocks for solid state qubits~\cite{uppu2021quantum, pezzagna2021quantum} or spin-photon interfaces~\cite{castelletto2020hexagonal}. These systems have the advantage that they couple efficiently to photons and can therefore be flawlessly interfaced with photonic infrastructures in the near-infrared to visible spectral range~\cite{lodahl2015interfacing}. However, a drawback of such an all-optical approach is the challenging miniaturization when approaching structure scales on the order of the operation wavelength of the light. When reaching this diffraction limit systems become increasingly lossy due to stray light~\cite{tan2015nonlinear} or internal absorption~\cite{xiong2021room}. Other approaches at even larger wavelengths rely on superconducting (transmon) qubits that operate with transition energies in the microwave range~\cite{houck2008controlling}. These operation energies naturally require very low operation temperatures below 1~K.\\
	To overcome the individual shortcomings, modern developments in quantum technology consider hybrid approaches that combine different solid state excitations to harness their combined strengths~\cite{xiang2013hybrid, kurizki2015quantum, schuetz2015universal, lachance2019hybrid, benito2020hybrid}, e.g., by relying on multiple quasi-particle excitations. In this work we consider phononic excitations as an additional common component for a hybrid infrastructure. As phonons are the basic excitations of crystals, they appear to be a natural choice in a solid state infrastructure because they remain confined to the system. Their slow propagation speed compared to photons can be used as an additional advantage because it reduces the diffraction limit significantly when operating at, e.g., microwave frequencies~\cite{delsing2019the}. Recent promising progress in this direction was made by coupling superconducting qubits to mechanical oscillators~\cite{lahaye2009nanomechanical, o2010quantum, mirhosseini2020superconducting} or acoustic waves~\cite{gustafsson2014propagating, chu2017quantum} which allowed to generate arbitrary phonon quantum states and therefore demonstrated full coherent quantum control in the microwave range~\cite{hofheinz2009synthesizing, satzinger2018quantum, chu2018creation, bild2022schr,schrinski2023macroscopic}. Still the connection between this new quantum acoustic direction and quantum optics in the visible spectral range renders a breakthrough goal for hybrid quantum infrastructures~\cite{choquer2022quantum}. So far the phonon-photon interface mediated by SPEs is only routinely implemented in the limit of a classical phonon field, e.g., in the form of surface acoustic waves (SAWs)~\cite{delsing2019the, nysten2020hybrid, wigger2021remote, imany2022quantum, decrescent2022large} or bulk waves~\cite{bruggemann2012laser,czerniuk2017picosecond}.\\
	In this work, we take a critical next step in the development of an acousto-optical quantum transducer between mechanical, i.e., phononic, and optical, i.e., photonic, quantum states. The center-piece of our theoretical approach is the independent boson model, which is a universal description that can be applied to different types of SPEs in the visible spectral range, e.g., quantum dots~\cite{krummheuer2002the} or color centers~\cite{duke1965phonon}, coupling to arbitrary phononic realizations, e.g., SAWs~\cite{weiss2021optomechanical} or mechanical resonators~\cite{wilson2004laser} and optical~\cite{groll2021controlling} or local phonon modes~\cite{wigger2019phonon}. In addition to these solid state realizations, the independent boson model, or suitable extensions to it, has been used in the description of molecules, e.g., to model the interaction between molecular plasmons and the molecule's vibrational modes~\cite{cui2016molecular}, to decribe the behavior of excitons in biomolecules inside a solvent~\cite{gilmore2005spin}, or to model the impact of vibrational modes on transport through molecular transistors~\cite{braig2003vibrational,chen2005effects}.
	
	In practice, we extend our recent results, based on generating functions, which we used to explore the transition regime between the classical and quantum acoustic interaction of an optically driven SPE~\cite{hahn2022photon}. With the approach presented in this work, based on the polaron frame instead of generating functions, we are able to derive analytic formulas that allow us to extract the phonon number statistics from the light-scattering spectrum of a SPE.
	
	After introducing the theory in Secs.~\ref{sec:model} and \ref{sec:RF_deriv}, we thoroughly characterize the main features of the resonance fluorescence (RF) spectra in Sec.~\ref{sec:properties} and finally introduce and analyze the read-out of the phonon number statistics for the example of an initial phonon state $(\ket{0}+\ket{1})/\sqrt{2}$ in Sec.~\ref{sec:reading}.
	
	\section{The optically driven independent boson model}\label{sec:model}
	Our model consists of a two-level system (TLS), coupled to a single phonon mode, where the transition energy of the TLS is significantly larger than the phonon energy. As potential realizations of this model, one could consider a quantum dot~\cite{stauber2000electron, krummheuer2002the, forstner2003phonon, hohenester2004quantum, nazir2016modelling, carmele2019non} or a color center~\cite{duke1965phonon, norambuena2016microscopic, wigger2019phonon, preuss2022resonant} (TLS) inside an acoustic or mechanical resonator~\cite{wilson2004laser} or coupled to a dispersion-less, e.g., optical, phonon mode~\cite{axt1999coherent, reiter2011generation, groll2021controlling}. The full Hamiltonian of our system, including optical driving, is given by~\cite{krummheuer2002the, alicki2004pure, roszak2006path, mahan2013many, nazir2016modelling}
	\begin{subequations}\label{eq:H}\begin{align}
		H(t)&=H_0+H_I(t)\,,\\
		H_0&=\hbar\omega_{\rm TLS} X^{\dagger}X+\hbar\omega b^{\dagger}b\notag\\
		&\qquad+X^{\dagger}X\left(\hbar gb^{\dagger}+\hbar g^*b^{}\right)\,,\\
		H_I(t)&=\frac{1}{2}\left[\mathcal{E}^*(t)X+\mathcal{E}(t)X^{\dagger}\right]\,,\label{eq:HI}
		\end{align}\end{subequations}
	where $X=\ket{G}\bra{X}$ is the transition operator from the excited state $\ket{X}$ to the ground state $\ket{G}$ of the TLS with transition frequency $\omega_{\rm TLS}$. $H_0$ is the independent boson model Hamiltonian, describing the coupling between the TLS and the phonon mode, the latter being modeled as a harmonic oscillator with frequency $\omega\ll\omega_{\rm TLS}$, annihilation (creation) operator $b^{(\dagger)}$ and coupling constant $g$. $H_I$ describes the interaction of the TLS with a classical, coherent light field in the usual dipole and rotating wave approximations, modeled by the function $\mathcal{E}(t)=-2\bra{X}\bm{d}\cdot\bm{E}(t)\ket{G}$. Here, $\bm{d}$ is the dipole operator and $\bm{E}$ is the positive-frequency part of the classical electric field of a laser.\\
	The Hamiltonian $H_0$ describes the system in absence of optical driving and can be diagonalized using the so-called polaron transformation with~\cite{mahan2013many, nazir2016modelling, alicki2004pure, roszak2006path}
	\begin{equation}
	T_P=\ket{G}\bra{G}+\ket{X}\bra{X}B_+\,,	
	\end{equation}
	and 
	\begin{equation}
	B_{\pm}=D(\pm \gamma)=\exp(\pm \gamma b^{\dagger} \mp \gamma^* b)
	\end{equation}
	being displacement operators for the phonon mode with coherent amplitude $\pm \gamma=\pm g/\omega$~\cite{glauber1963coherent}. They have the properties $B_\pm^{\dagger}=B_\mp$ and $B_\pm B_\mp=1$, rendering the polaron transformation unitary. With this transformation, $H_0$ can be diagonalized, leading to
	\begin{equation}\label{eq:H0_pf}
	T_P^{}H_0T_P^{\dagger}=\hbar\omega_{\rm ZPL} X^{\dagger}X +\hbar\omega b^{\dagger}b\,,
	\end{equation}
	where
	\begin{equation}
	\omega_{\rm ZPL}=\omega_{\rm TLS}-\frac{|g|^2}{\omega}=\omega_{\rm TLS}-\omega|\gamma|^2
	\end{equation}
	is the polaron-shifted TLS frequency, i.e., the frequency of the zero-phonon line (ZPL). The eigenstates of $H_0$ in the polaron frame, i.e., the eigenstates of $T_P^{}H_0 T_P^{\dagger}$, are given by the product states $\ket{G}\ket{n}$ and $\ket{X}\ket{n}$ with $\ket{n}$ denoting the $n$-phonon Fock state. Their energies are $n\hbar\omega$ and $\hbar\omega_{\rm ZPL}+n\hbar\omega$, respectively. The eigenstates of $H_0$ (in the original frame) are thus given by
	\begin{subequations}
		\begin{align}
		\ket{n}_G&=T_P^{\dagger}\ket{G}\ket{n}=\ket{G}\ket{n}\,,\\
		\ket{n}_X&=T_P^{\dagger}\ket{X}\ket{n}=\ket{X}B_-\ket{n}\label{eq:nx}
		\end{align}
	\end{subequations}
	and the corresponding eigenvalue equations read
	\begin{subequations}\label{eq:H0_spec}\begin{align}
		H_0\ket{n}_G&=n\hbar\omega \ket{n}_G\,,\\
		H_0\ket{n}_X&=\left(\hbar \omega_{\rm ZPL}+n\hbar\omega \right)\ket{n}_X\,.\label{eq:H0_spec_x}
		\end{align}\end{subequations}
	\begin{figure}[t]
		\centering
		\includegraphics[width=0.4\linewidth]{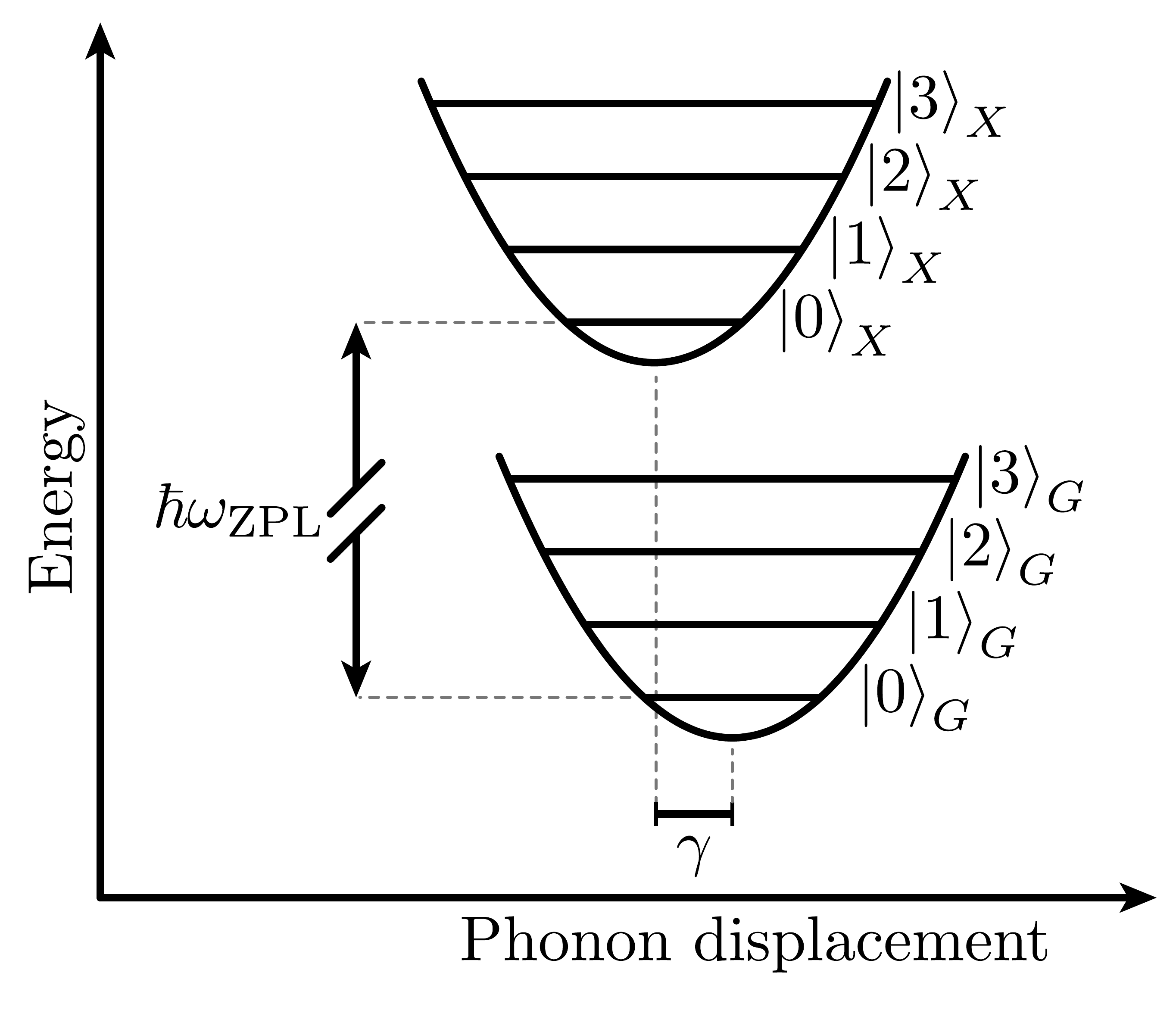}
		\caption{Schematic representation of the spectrum of $\textit{H}_0$ in Eqs.~\eqref{eq:H0_spec}. It consists of two harmonic oscillators, energetically shifted relative to each other by the ZPL energy $\hbar\omega_\text{ZPL}$. The harmonic oscillators are separated in phase space by the displacement given in terms of the dimensionless phonon coupling constant $\gamma$.}
		\label{fig:schema1}
	\end{figure}\noindent
	The spectrum of $H_0$ therefore consists of two harmonic oscillator spectra, shifted relative to each other by the ZPL energy $\hbar\omega_{\rm ZPL}$, as displayed in Fig.~\ref{fig:schema1}. The harmonic oscillator associated with $\ket{X}$ is displaced relative to the harmonic oscillator associated with $\ket{G}$ by the dimensionless coupling strength $\gamma$. This is mathematically described by the appearance of the displacement operator $B_-$ in Eq.~\eqref{eq:nx} and is due to the shift in the equilibrium positions of the lattice atoms originating from the change in charge distribution when exciting the TLS~\cite{mahan2013many, hohenester2007quantum, wigger2014energy}.\\
	The full Hamiltonian in the polaron frame $T_P^{}H(t)T_P^{\dagger}$ is obtained by adding Eq.~\eqref{eq:H0_pf} and the optical driving
	\begin{equation}\label{eq:HI_pf}
	T_P^{}H_I(t)T_P^{\dagger}=\frac{1}{2}\qty[\mathcal{E}^*(t)X B_-+\mathcal{E}(t) X^{\dagger} B_+]\,.
	\end{equation}
	Whereas in the original frame the optical driving in Eq.~\eqref{eq:HI} did not involve any phonon operators, in the polaron frame the displacement operators $B_{\pm}$ appear. This implies that the optical driving couples the polaron and the phonons in such a way that transitions between different phonon states are possible when exciting/de-exciting the polaron. The non-vanishing matrix elements between eigenstates of $H_0$ in the polaron frame, i.e., eigenstates of $T_P^{}H_0T_P^{\dagger}$, are given by
	\begin{equation}\label{eq:def_FC}
	\bra{X}\bra{m} T_P^{} H_I(t) T_P^{\dagger}\ket{n}\ket{G}=\braind{X}{m}H_I(t)\ket{n}_G=\frac{1}{2}\mathcal{E}(t) \bra{m}B_+\ket{n}=\frac{1}{2}\mathcal{E}(t) M_m^n
	\end{equation}
	and their complex conjugate. Here, $\bra{m}B_+\ket{n}=M_{m}^n$ denote the Franck-Condon (FC) factors~\cite{franck1926elementary, condon1926theory, maguire2019environmental}. These are overlap integrals between the phonon state $\ket{n}$ associated with the TLS ground state and the phonon state $B_-\ket{m}$ associated with the TLS excited state and determine the probability for a corresponding optical transition.\\
	In addition to the Hamiltonian in Eqs.~\eqref{eq:H} we include two Lindblad dissipators describing pure dephasing (pd) and excited state decay (xd) of the TLS with the rates $\gamma_{\rm pd}$ and $\gamma_{\rm xd}$, respectively, given by~\cite{breuer2002theory, groll2021controlling}
	\begin{subequations}\begin{align}
		\mathcal{D}_{\rm pd}\left(\rho \right)&=\gamma_{\rm pd}\left(X^{\dagger}X\rho X^{\dagger}X-\frac{1}{2}\left\lbrace X^{\dagger}X,\rho\right\rbrace\right)\,,\label{eq:D_pd}\\
		\mathcal{D}_{\rm xd}\left(\rho \right)&=\gamma_{\rm xd}\left(X\rho X^{\dagger}-\frac{1}{2}\left\lbrace X^{\dagger}X,\rho\right\rbrace\right)\,.\label{eq:D_xd}
		\end{align}\end{subequations}
	This extension of our model is necessary since every physical realization of an excitonic TLS, e.g., a color center, is subject to dissipation, which has to be included in realistic modeling of experimental observables. The full time evolution of the density matrix $\rho$, describing the state of the coupled TLS-phonon system, is finally given by
	\begin{equation}\label{eq:t-evol}
	\frac{\text{d}}{\text{d}t}\rho(t)=-\frac{i}{\hbar}\left[H(t),\rho(t)\right]+\mathcal{D}_{\mathrm{pd}}\left[\rho(t)\right]+\mathcal{D}_{\mathrm{xd}}\left[\rho(t)\right]\,.
	\end{equation}
	\section{Resonance fluorescence signal}\label{sec:RF_deriv}
	\begin{figure}[h]
		\centering
		\includegraphics[width=0.4\linewidth]{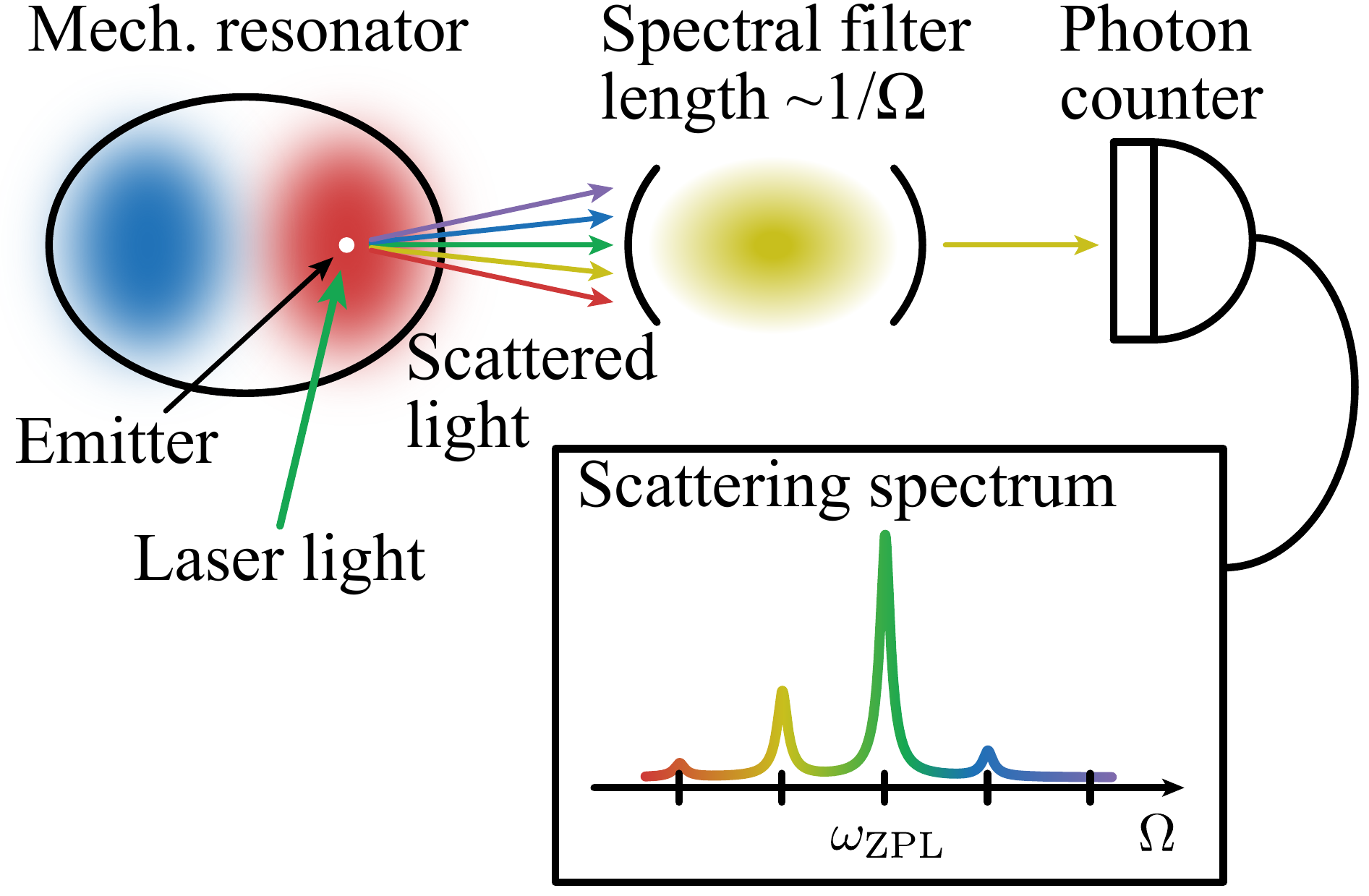}
		\caption{Schematic representation of the RF experiment. An emitter coupled to a phonon mode of a mechanical resonator is driven by laser light. The emitted light is filtered by an interferometer and then detected by a photon counter. In this prototypical spectrometer, the length of the interferometer determines the detection frequency $\Omega$.}
		\label{fig:schema2}
	\end{figure}\noindent
	In this section, we derive an analytic formula for the RF signal in the driven independent boson model for arbitrary initial phonon states. This will be used later to establish a connection between the number statistics of the initial phonon state and the form of the RF spectrum. We consider continuous-wave (cw) excitation and investigate the spectral components of the light that is scattered from the TLS, as displayed in Fig.~\ref{fig:schema2}. Before moving on with the derivation of the RF signal, however, we will shortly discuss the appropriate interpretation of the experimental situation that is modeled here.\\
	Any physical realization of our model will contain a finite phonon lifetime $T_\mathrm{ph}$. This was implicitly chosen to be $T_\mathrm{ph}\rightarrow\infty$ in Sec.~\ref{sec:model}, assuming that it is significantly longer than $\gamma_{\rm pd}^{-1}$ and $\gamma_{\rm xd}^{-1}$. If we were driving the system (including a finite phonon lifetime) for an infinite amount of time, collecting all scattered photons, we would inevitably lose the information on the initial phonon state. The reason is, that in a typical periodically-driven dissipative system the information on the initial preparation is lost over time~\cite{kohler1997floquet,lazarides2014equilibrium}.	This argument makes it clear that we cannot be interested in a true stationary signal here, but we rather want to investigate the transient of the phonon subsystem, which still contains the information on the initial phonon state. To be more specific, the timescale of detection $T$ is assumed to be much smaller than the phonon lifetime $T\ll T_\mathrm{ph}$ while being much larger than the phonon period or the lifetime of the TLS $T\gg \omega^{-1},\gamma_{\rm xd}^{-1}$. These constraints require a sufficiently high quality factor for the considered phonon mode. Potential realizations could be mechanical resonators formed by suspended monolayer transition metal dichalcogenides~\cite{morell2016high} containing localized excitons~\cite{tonndorf2015single}, carbon nanotube mechanical resonators~\cite{huttel2009carbon,laird2012high} with optically active defects~\cite{he2017tunable} or quantum dots in SAW resonators~\cite{nysten2020hybrid}. All of these implementations have quality factors of $Q_{\rm ph}\approx T_{\rm ph}\omega \gtrsim 10^4$, which opens a wide range of easily accessible optical measurement times $T$.\\
	Furthermore, we consider weak optical driving in the following, such that on the detection timescale $T$ only a single photon scattering process takes place. This implies that the Rabi frequency of the cw-laser has to be much smaller than $T^{-1}$, which yields another upper bound on the detection time $T$. In the following calculations this is secured by considering only those contributions to the RF signal, that are of second order in the driving field $\mathcal E$. We will formally take the limit $T\rightarrow\infty$, keeping in mind that $T$ is in practice still limited by the phonon lifetime and the inverse of the Rabi frequency of the cw-laser.\\
	It is clear that a single scattered and detected photon is not sufficient to measure a complete RF spectrum. Thus, we have to interpret the derived RF signal as stemming from an average of repeated measurements on identically prepared TLS's with the same initial phonon state. To conclude, we seek for the connection between the statistics of the scattered light, i.e., the RF signal, and the number statistics of the initial phonon state during the transient of the phonon subsystem.\\	
	The starting point for the derivation of the RF signal is the time-integrated spectrum, detected by a prototypical detector comprised of a Fabry-Perot interferometer (spectral filter) with resolution $\Gamma$ and setting-frequency $\Omega$ and a photon counter, as displayed in Fig.~\ref{fig:schema2}~\cite{eberly1977time}. Note, that the derivation of the final RF signal is independent of the particular choice of the detector model and one could in principle also choose other detector response functions than the one used here~\cite{preuss2022resonant}. A derivation of the time-integrated spectrum, starting from the time-dependent spectrum~\cite{eberly1977time}, is given in Sec.~S1 of the Supporting Information (SI). The time-integrated spectrum reads
	\begin{equation}\label{eq:S_RF}
	S(\Omega;\Gamma)=\Gamma\text{Re}\left[\int\limits_0^{\infty}\text{d}\tau\,e^{-(\Gamma+i\Omega)\tau}\overline{G}(\tau)\right]\,,
	\end{equation}
	where $\overline{G}$ is the time-averaged correlation function
	\begin{equation}\label{eq:G_bar}
	\overline{G}(\tau)=\lim\limits_{T\rightarrow\infty}\frac{1}{T}\int\limits_{t_0}^{t_0+T}\text{d}t\,G(t+\tau,t)
	\end{equation}
	with~\cite{mollow1969power}
	\begin{align}\label{eq:G_t_tau}
	G(t+\tau,t)&=\expval{X^{\dagger}(t+\tau)X(t)}\\
	&=\text{Tr}\Big(X^{\dagger}\mathcal{V}(t+\tau,t)\left\lbrace X\mathcal{V}(t,t_0)\left[\rho(t_0)\right]\right\rbrace\Big)\,.\notag
	\end{align}
	We start with the initial density matrix in the form
	\begin{equation}
	\rho(t_0)=\ket{G}\bra{G}\rho^G\,,
	\end{equation}
	where $\rho^G$ is an arbitrary initial phonon state associated with the TLS's ground state. In Eq.~\eqref{eq:G_t_tau} we apply the quantum regression theorem with respect to the Markovian dissipators $\mathcal{D}_\mathrm{pd}$ and $\mathcal{D}_\mathrm{xd}$, while the phonons are an explicit part of our system's Hamiltonian. The non-unitary time evolution super-operator $\mathcal{V}(t,t_0)$ used therein has the property~\cite{breuer2002theory}
	\begin{equation}
	\mathcal{V}(t,t_0)\left[\rho(t_0)\right]=\rho(t)\,.
	\end{equation}
	When inserting $1=\ket{G}\bra{G}+\ket{X}\bra{X}$ to the right of $\rho(t)=\mathcal{V}(t,t_0)[\rho(t_0)]$ in Eq.~\eqref{eq:G_t_tau}, we see that there are two contributions to the correlation function $G(t+\tau,t)$~\cite{hahn2022photon}. These are determined by the parts of the density matrix $\rho(t)$ in Eq.~\eqref{eq:G_t_tau}, which do not vanish after multiplication by $X=\ket{G}\bra{X}$ from the left, namely $\bra{X}\rho(t)\ket{X}$ and $\bra{X}\rho(t)\ket{G}$, which are explicitly derived in Sec.~S2.1 and S2.2 of the SI, respectively. To determine these contributions, we start by separating the full time evolution in Eq.~\eqref{eq:t-evol} into a free part and an interaction part, including the optical field, via
	\begin{subequations}\begin{align}
		\frac{\text{d}}{\text{d}t}\rho(t)&=\mathcal{L}_0\qty[\rho(t)]+\mathcal{L}_I(t)\qty[\rho(t)]\,,\\
		\mathcal{L}_0\qty[\rho(t)]&=-\frac{i}{\hbar}\left[H_0,\rho(t)\right]+\mathcal{D}_{\mathrm{pd}}\qty[\rho(t)]+\mathcal{D}_{\mathrm{xd}}\qty[\rho(t)]\,, \label{eq:free}\\
		\mathcal{L}_I\qty[\rho(t)]&=-\frac{i}{\hbar}\left[H_I(t),\rho(t)\right]\,.
		\end{align}\end{subequations}
	We are interested in the case of weak optical driving and in the following perform a perturbation expansion in orders of $H_I$. Up to second order in $H_I$, the density matrix at time $t$ is given by
	\begin{align}\label{eq:expansion}
	\rho(t)&=e^{\mathcal{L}_0(t-t_0)}\rho(t_0) \notag\\
	&+\int\limits_{t_0}^t\text{d}\tau_1\,e^{\mathcal{L}_0(t-\tau_1)}\mathcal{L}_I(\tau_1)e^{\mathcal{L}_0(\tau_1-t_0)}\rho(t_0)\notag\\
	&+\int\limits_{t_0}^t\text{d}\tau_1\,\int\limits_{t_0}^{\tau_1}\text{d}\tau_2\,e^{\mathcal{L}_0(t-\tau_1)}\mathcal{L}_I(\tau_1)e^{\mathcal{L}_0(\tau_1-\tau_2)} \notag\\
	&\qquad \qquad \qquad \times\mathcal{L}_I(\tau_2)e^{\mathcal{L}_0(\tau_2-t_0)}\rho(t_0)\,.
	\end{align}
	Here, we dropped the rectangular brackets, e.g.,  $\exp(\mathcal{L}\tau)\qty[\rho(t)]\rightarrow \exp(\mathcal{L}\tau)\rho(t)$,  and use the convention that the superoperators $\mathcal{L}_0$ and $\mathcal{L}_I$, as well as their exponentials, simply act on everything to their right.\\
	To model the typical RF experiment~\cite{mollow1969power, weiss2021optomechanical, wigger2021resonance}, which uses a cw-laser, switched on at time $t=t_0$, to drive the TLS, we choose
	\begin{equation}
	\mathcal{E}(t)=\mathcal{E}_0e^{-i\omega_lt}\Theta(t-t_0)\,,
	\end{equation}
	where $\omega_l=\omega_{\rm ZPL}+\kappa\omega$ is the carrier frequency of the laser and the detuning relative to the ZPL is given by $\delta=\omega_l-\omega_{\rm ZPL}=\kappa\omega$ with an arbitrary $\kappa \in \mathbb R$.\\
	A detailed derivation of the time-averaged correlation function is presented in Sec.~S2 of the SI. A key approximation therein is the neglect of any transient of the TLS, i.e., $t-t_0\gg \gamma_{\rm xd}^{-1}$ in Eq.~\eqref{eq:G_t_tau} and we assume $t-t_0\gg \omega^{-1}$. Both of these assumptions can be realized by using a large enough detection time $T$ in Eq.~\eqref{eq:G_bar}, such that the relative weight of the transient dynamics becomes negligible.\\
	The density matrix in the excited state subspace, which is calculated in Sec.~S2.1 of the SI and of which we will make use in the following discussions, is given by
	\begin{align}
	&\braind{X}{m}\rho(t)\ket{n}_X =\sum_{p,q}M_m^pM_n^{q*}\rho_{p,q}^Ge^{-i\omega(p-q)(t-t_0)}\notag\\
	&\qquad\times\frac{|\mathcal{E}_0|^2}{4\hbar^2}\frac{\gamma_{\rm pd}+\gamma_{\rm xd}+i\omega(m-n-p+q)}{\gamma_{\rm xd}+i\omega(m-n-p+q)}\notag\\
	&\qquad\times\left[\frac{\gamma_{\rm pd}+\gamma_{\rm xd}}{2}+i\omega(\kappa+q-n)\right]^{-1}\notag\\
	&\qquad\times\left[\frac{\gamma_{\rm pd}+\gamma_{\rm xd}}{2}-i\omega(\kappa+p-m)\right]^{-1}\!.\label{eq:rho_X}
	\end{align}
	Here, the FC factors $\bra{m}B_+\ket{n}=M_{m}^n$ from Eq.~\eqref{eq:def_FC} appear. We have also introduced the shorthand notation $\bra{p}\rho^G\ket{q}=\rho_{p,q}^G$ for the initial phonon density matrix elements. We see that the density matrix elements associated with the excited state of the TLS $\braind{X}{m}\rho(t)\ket{n}_X$ are periodic with respect to the phonon frequency $\omega$ but otherwise time-independent. This shows that the TLS approaches a stationary state for $t-t_0\gg\gamma_{\rm xd}^{-1}$ in second order with respect to the driving field, where the weak optical driving and the dissipation are balanced.\\
	The final time-averaged correlation function, for which we also need to take into account contributions from the TLS coherence $\bra{X}\rho(t)\ket{G}$, is given by
	\begin{align}
	&\overline{G}(\tau)= \notag\\
	&\frac{|\mathcal{E}_0|^2}{4\hbar^2}\sum_{m,n,q,p} \rho_{p,p}^GM_n^m M_q^{m*}M_q^pM_n^{p*}e^{i\omega_{\rm ZPL}\tau}e^{-i(m-n)\omega\tau}\notag\\
	&\quad\qquad \times \left[\frac{\gamma_{\rm pd}+\gamma_{\rm xd}}{2}+i\omega(\kappa-n+p)\right]^{-1}\notag\\
	&\quad\qquad\times\left[\frac{\gamma_{\rm pd}+\gamma_{\rm xd}}{2}-i\omega(\kappa-q+p)\right]^{-1}\notag\\
	&\quad\qquad\times\left[e^{i(\kappa+p-n)\omega \tau}+\frac{\gamma_{\rm pd}e^{-(\gamma_{\rm pd}+\gamma_{\rm xd})\tau/2}}{\gamma_{\rm xd}+i\omega(q-n)}\right]\!.\label{eq:average_G_total}
	\end{align}
	We see that the time-averaged correlation function and thus the RF signal only depend on the phonon occupations $\rho_{p,p}^G$, i.e., the phonon number statistics, due to the temporal averaging. This means that coherences of the initial phonon quantum state, i.e., off-diagonal elements of the initial density matrix, do not contribute here. This limitation could be overcome by considering time-dependent instead of time-integrated spectroscopy signals~\cite{groll2021controlling,wigger2021resonance} which is, however, beyond the scope of the present paper. Also, there are two contributions: One that oscillates in $\tau$ and one that additionally decays with the rate $(\gamma_{\rm pd}+\gamma_{\rm xd})/2$ [see last line in Eq.~\eqref{eq:average_G_total}]. However, the latter vanishes in the limit $\gamma_{\rm pd}\rightarrow 0$. It constitutes a background which is only present in the case of non-vanishing pure dephasing, stemming from inelastic scattering as will be explained in detail in Sec.~\ref{sec:bkg}. This background was already found numerically in Ref.~\cite{wigger2021remote} and is here derived rigorously. Plugging this full expression for $\overline{G}(\tau)$ into Eq.~\eqref{eq:S_RF}, we obtain the RF spectrum
	\begin{equation}
	S(\Omega;\Gamma)=\sum_{N=-\infty}^{\infty} \left[S^{N+\kappa}_{\rm es}(\Omega;\Gamma)+S^N_{\rm is}(\Omega;\Gamma)\right]\,,\label{eq:S_total_complicated}
	\end{equation}
	which we separate into elastic scattering (es) contributions $S^{N+\kappa}_{\rm es}$ at frequencies $\Omega=\omega_{\rm ZPL}+N\omega+\kappa\omega$ according to the restriction $N=p-m$ under the sum in Eq.~\eqref{eq:average_G_total} and inelastic scattering (is) contributions $S^N_{\rm is}$ around frequencies $\Omega=\omega_{\rm ZPL}+N\omega$ according to the restriction $N=n-m$. These contributions read
		\begin{subequations}\begin{align}
			S^{N+\kappa}_{\rm es}(\Omega;\Gamma) &= \frac{|\mathcal{E}_0|^2}{4\hbar^2}\underset{p-m=N}{\sum_{m,n,q,p}} \rho_{p,p}^GM_n^m M_q^{m*}M_q^pM_n^{p*}
			\frac{\Gamma^2}{\Gamma^2+\left[\Omega-\omega_{\rm ZPL}-(N+\kappa)\omega\right]^2} \notag\\
			&\qquad\times \left[\frac{\gamma_{\rm pd}+\gamma_{\rm xd}}{2}+i\omega(\kappa-n+p)\right]^{-1} 
			\left[\frac{\gamma_{\rm pd}+\gamma_{\rm xd}}{2}-i\omega(\kappa-q+p)\right]^{-1}\!,\label{eq:S_peak} \\
			S^N_{\rm is}(\Omega;\Gamma) &= 
			\frac{\text{Re}(\phi_N)\Gamma\left(\Gamma+\frac{\gamma_{\rm pd}+\gamma_{\rm xd}}{2}\right) + \text{Im}(\phi_N)\Gamma(\Omega-\omega_{\rm ZPL}-N\omega)}{\left(\Gamma+\frac{\gamma_{\rm pd}+\gamma_{\rm xd}}{2}\right)^2+\left(\Omega-\omega_{\rm ZPL}-N\omega\right)^2} \,,\label{eq:S_bkg} \\
			\text{with}\quad \phi_N&= 
			\frac{ |\mathcal{E}_0|^2}{4\hbar^2} \underset{N=n-m}{\sum_{m,n,q,p}}\rho_{p,p}^GM_{n}^mM_q^{m*}M_q^pM_{n}^{p*}\frac{\gamma_{\rm pd}}{\gamma_{\rm xd}+i\omega(q-n)}\notag\\
			&\qquad\times\left[\frac{\gamma_{\rm pd}+\gamma_{\rm xd}}{2}+i\omega(\kappa-n+p)\right]^{-1}
			\left[\frac{\gamma_{\rm pd}+\gamma_{\rm xd}}{2}-i\omega(\kappa-q+p)\right]^{-1}\!.\label{eq:phi_N}
			\end{align}\end{subequations}
	The physical content of these rather lengthy expressions will be discussed in Sec.~\ref{sec:properties}, as well as typical limiting cases in which the form of the RF spectrum simplifies significantly. We can already note two things: (A) the spectrum consists of sharp Lorentzian peaks at frequencies $\Omega=\omega_{\rm ZPL}+N\omega+\kappa\omega$, broadened only by the resolution $\Gamma$ of the spectrometer. This reflects, that in RF the resolution is only limited by the spectrometer and the spectral width of the laser~\cite{wu1975investigation}. Since we performed the calculation for an ideal cw-laser, whose spectral width is zero, at ideal spectrometer resolution $\Gamma\rightarrow 0$ the spectrum becomes a set of $\delta$-peaks. These peaks appear at multiple phonon frequencies relative to the laser frequency $\omega_{\rm ZPL}+\kappa\omega$. (B) For non-vanishing pure dephasing, there are inelastic contributions around multiple phonon frequencies relative to the ZPL $\Omega=\omega_{\rm ZPL}+N\omega$ with their spectral positions being independent of the laser frequency. These are sums of two lineshapes: (i) A broad, symmetric Lorentzian with the width $\Gamma+\frac{\gamma_{\rm pd}+\gamma_{\rm xd}}{2}$ determined by the spectrometer resolution and the TLS dissipation rates, whose weight is determined by the real part of the function $\phi_N$. (ii) An antisymmetric lineshape of the same width as (i), whose weight is determined by the imaginary part of $\phi_N$. Thus, in general the inelastic background around $\Omega=\omega_{\rm ZPL}+N\omega$ is not symmetric.
	\section{General properties and limiting cases of the RF spectrum}\label{sec:properties}
	\subsection{Qualitative derivation of the RF spectrum for vanishing pure dephasing}\label{sec:RF_qual_deriv}
	\begin{figure}[h]
		\centering
		\includegraphics[width=0.5\columnwidth]{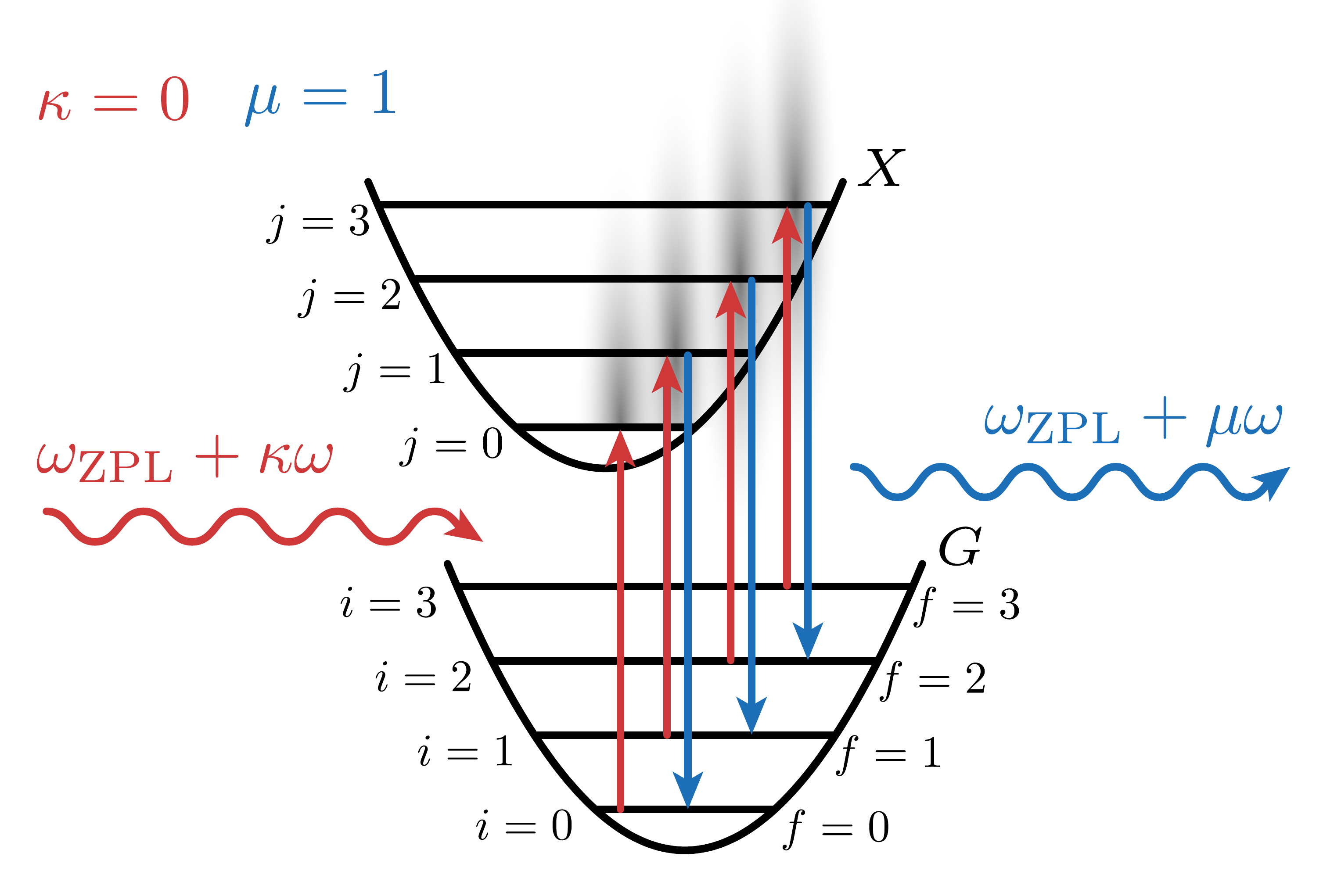}
		\caption{Representation of the transitions from possible initial states ($i$) to intermediate states ($j$) induced by laser light of frequency $\omega_\text{ZPL}+\kappa\omega$ (red arrows) for $\kappa=0$ (ZPL excitation). Subsequent transitions to final states ($f$) leads to the emission of light with frequency $\omega_\text{ZPL}+\mu\omega$ with $\mu=\text{1}$ here (blue arrows). The intermediate resonances are broadened due to finite dissipation rates $\gamma_\text{pd}$ and $\gamma_\text{xd}$, as indicated by the grey shaded areas.}
		\label{fig:schema3}
	\end{figure}
	We will discuss the limiting case of $\gamma_{\rm pd}=0$ first and present a qualitative derivation of the final results from the previous section, which will clarify much better the underlying physics. In this limit, there is no broad background and the emitted photons have well-defined frequencies. The physical process that leads to the RF spectrum is illustrated in Fig.~\ref{fig:schema3} and can be explained as follows: A photon of frequency $\omega_{\rm ZPL}+\kappa\omega$ scatters from the system, which is in an initial (phonon) state $\ket{i}_G$ and performs a transition to some final (phonon) state $\ket{f}_G$ (red arrows followed by blue arrows). In the limit of $(t-t_0)\gg \gamma_{\rm xd}^{-1}$, energy conservation restricts the photon's frequency after the scattering to $\omega_{\rm ZPL}+\kappa\omega-(f-i)\omega$. Consequently, only this frequency is measured. Thus, to get a measure for the amplitude of the peak at frequency $\omega_{\rm ZPL}+\mu\omega$, we have to sum over all possible processes leading to a frequency of $\mu\omega=(\kappa-f+i)\omega$. Since the final state is not directly measured, we also have to take the average over all possible final states. As the initial state is in general not a pure state, when summing over all initial states, we have to weight each possibility with the appropriate probabilities, given by the density matrix occupations $\rho_{i,i}^G$, as mentioned before. Coherences do not contribute since they average out in the time integration, as discussed previously. Furthermore, we have to sum over all possible intermediate excited states $\ket{j}_X$ weighted by the transition amplitudes~\cite{feynman2010quantum}
	\begin{align}
	\braind{G}{f}&X\ket{j}\!{}^{}_X \braind{X}{j}X^{\dagger}\ket{i}_G\notag\\
	&=\bra{f}B_-\ket{j}\bra{j}B_+\ket{i}=M_j^iM_j^{f*}
	\end{align}
	for their excitation and de-excitation. In the limit of $\gamma_{\rm xd}\rightarrow 0$, energy conservation would impose the restriction of $j=i+\kappa$. This selection rule is of course weakened by the broadening of the resonances due to radiative decay (grey shaded areas in Fig.~\ref{fig:schema3}), leading to an overall scattering amplitude of
	\begin{align}
	&\ket{i}_G\rightarrow\ket{f}_G\text{ via }\ket{j}_X\notag\\
	&\sim M_j^i M_j^{f*}\left[\frac{\gamma_{\rm xd}}{2}-i\omega(i+\kappa-j)\right]^{-1}\!.
	\end{align}
	Summing over all intermediate states, averaging over the final and initial states, the probability $p_{\mu}$ to measure a photon of frequency $\omega_{\rm ZPL}+\mu\omega$ is finally given by
	\begin{equation}
	p_{\mu}\sim \!\!\!\!\! \underset{\mu=\kappa-f+i}{\sum_{i,f}} \!\!\!\! \rho_{i,i}^G\left|\sum_{j}M_j^i M_j^{f*}\left[\frac{\gamma_{\rm xd}}{2}-i\omega(i+\kappa-j)\right]^{-1}\right|^2\!.
	\end{equation}
	In total, this implies a RF spectrum of the form
	\begin{align}
	&S(\Omega;\Gamma)\notag\\
	&\sim\sum_{\mu}  \frac{p_{\mu}}{\Gamma^2+\left(\Omega-\omega_{\rm ZPL}-\mu\omega\right)^2}\notag\\
	&\sim \sum_{i,f,j_1,j_2} \frac{\rho_{i,i}^G M_{j_1}^iM_{j_1}^{f*}M_{j_2}^{i*}M_{j_2}^{f}}{\Gamma^2+\left[\Omega-\omega_{\rm ZPL}-(\kappa-f+i)\omega\right]^2}\label{eq:spec_paths} \\
	&\times\left[\frac{\gamma_{\rm xd}}{2}-i\omega(i+\kappa-j_1)\right]^{-1}\!\left[\frac{\gamma_{\rm xd}}{2}+i\omega(i+\kappa-j_2)\right]^{-1}\!\!. \notag
	\end{align}
	This is exactly the same result as obtained when summing over the peak contributions from Eq.~\eqref{eq:S_peak} with the replacements $p\rightarrow i,\ m\rightarrow f,\ q\rightarrow j_1,\ n\rightarrow j_2$. Thus, the peak at $\Omega=\omega_{\rm ZPL}+N\omega+\kappa\omega$ stems from scattering of photons which induce transitions of the form $\ket{i}_G\rightarrow\ket{f}_G=\ket{i-N}_G$ in the phonon state via some unknown intermediate excited states.\\
	In the case of resonant excitation on a phonon sideband (PSB) or the ZPL, i.e., $\kappa\in\mathbb{Z}$, and a slow decay with $\gamma_{\rm xd}\ll\omega$, we can make use of the approximation 
	\begin{equation}
	\left[\frac{\gamma_{\rm xd}}{2}+i\omega(\kappa-k)\right]^{-1}\approx \frac{2}{\gamma_{\rm xd}}\delta_{\kappa,k}\,.
	\end{equation}
	In this limit, only strictly resonant processes contribute to the spectrum, which simplifies Eq.~\eqref{eq:spec_paths} substantially, leading to
	\begin{equation}
	S(\Omega;\Gamma)\sim \sum_{i,f}\frac{\rho_{i,i}^G |M_{i+\kappa}^iM_{i+\kappa}^{f}|^2}{\Gamma^2+\left[\Omega-\omega_{\rm ZPL}-(\kappa-f+i)\omega\right]^2}\,.
	\end{equation}
	\subsection{Discussion of the inelastic contributions}\label{sec:bkg}
	For a non-vanishing pure dephasing rate $\gamma_{\rm pd}\neq 0$ additional background contributions appear in the RF spectrum [see Eq.~\eqref{eq:S_bkg}]~\cite{wigger2021remote}. Using the replacements $p\rightarrow i,\ m\rightarrow f,\ q\rightarrow j_1,\ n\rightarrow j_2$ from the previous section, which give physical meaning to the indices, we see that the background at $\Omega=\omega_{\rm ZPL}+N\omega$ is due to processes with $N=j_2-f$ involving an intermediate excited state $\ket{j_2}_X$ and the final phonon state $\ket{f}_G$. This can be interpreted as an inelastic scattering process in which the incoming photon leads to a transition $\ket{i}_G\rightarrow\ket{j_2}_X$ even for $\kappa\omega\neq (j_2-i)\omega$ and after a while a photon of frequency $\omega_{\rm ZPL}+(j_2-f)\omega$ is spontaneously emitted. The latter part of this process is effectively included in our model via the dissipator $\mathcal{D}_\mathrm{xd}$ in Eq.~\eqref{eq:D_xd}, while the first part will be discussed in the following. This interpretation is emphasized by the fact that the position of each inelastic background contribution $S^N_{\rm is}(\Omega)$ does not depend on the detuning $\kappa\omega$ of the laser (where $\kappa\in\mathbb{R}$ in general). Its shape is of course influenced by the $\phi_N$ from Eq.~\eqref{eq:phi_N}, but the spectral position is always at $\Omega=\omega_{\rm ZPL}+N\omega$ with $N\in\mathbb{Z}$, implying that the incoming photon is completely absorbed and the subsequent emission of light is due to a decay process $\ket{j_2}_X\rightarrow\ket{f}_G$.\\
	\begin{figure}[h]
		\centering
		\includegraphics[width=0.5\linewidth]{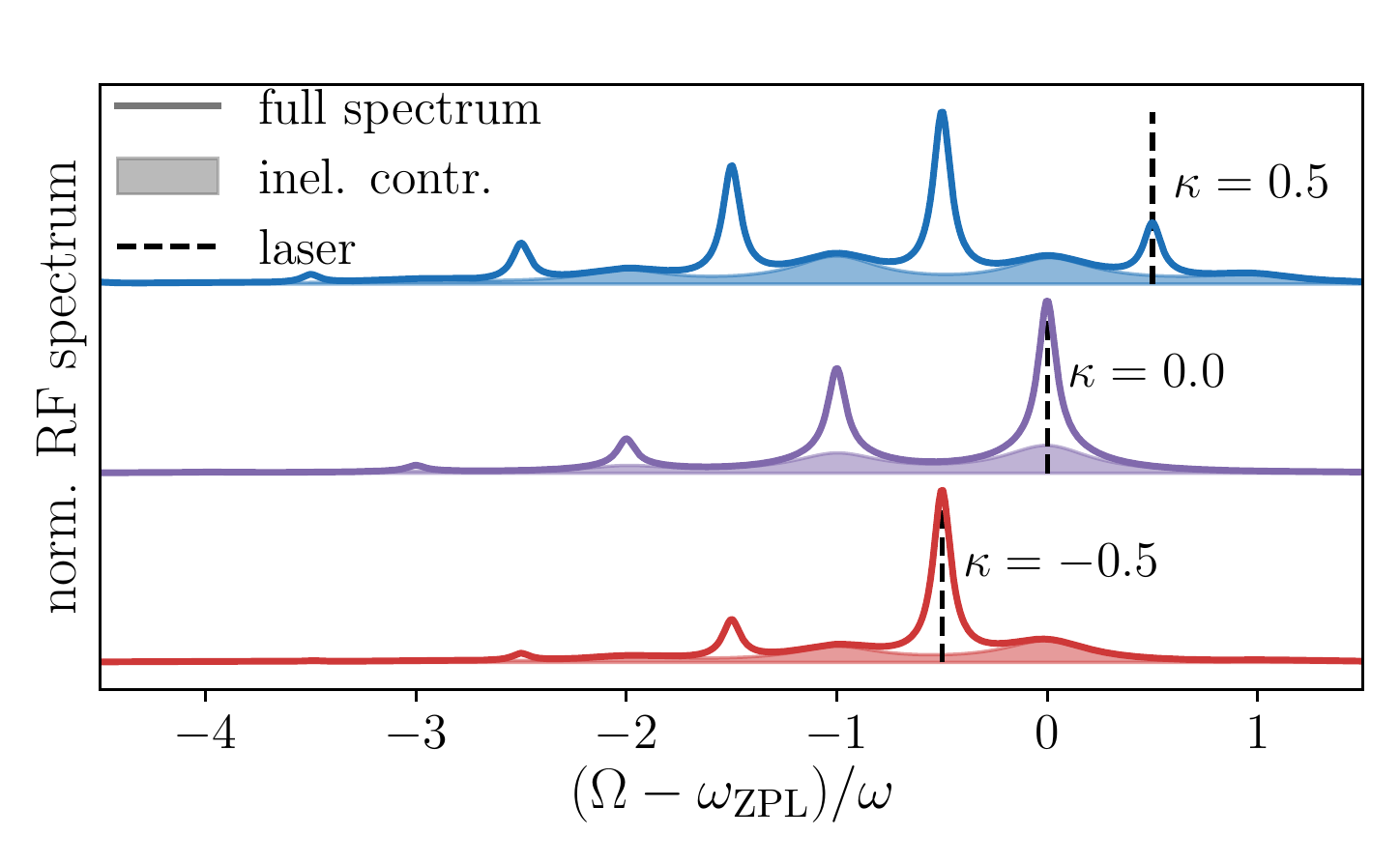}
		\caption{Examples of RF spectra (colored lines) for the case of an initial phononic vacuum $\rho^\textit{G}=\ket{0}\bra{0}$, employing the full formula from Eq.~\eqref{eq:S_total_complicated}. The spectrometer resolution is $\Gamma=\text{0.05}\omega$ and the dissipation rates are chosen as $\gamma_\text{pd}=\gamma_\text{xd}=\text{0.2}\omega$. The phonon coupling is $\gamma=\text{0.8}\omega$. The background due to inelastic scattering (colored shaded areas) is spectrally separated from the elastic contributions by varying the laser detuning $\kappa\omega$ as indicated by the vertical dashed lines. The values for $\kappa$ are given in the plot.}
		\label{fig:bkg}
	\end{figure}
	Figure~\ref{fig:bkg} displays the discussed features. It shows RF spectra (colored lines) for the case of an initial phononic vacuum $\rho^G=\ket{0}\bra{0}$. The other relevant parameters are $\Gamma=0.05\omega$, $\gamma_{\rm pd/xd}=0.2\omega$ and $\gamma=0.8\omega$. The spectra are displayed for different detunings $\kappa\omega$, considering excitation slightly above/below the ZPL ($\kappa=\pm0.5$, blue/red line) and resonant excitation on the ZPL ($\kappa=0$, purple line). In the latter case the spectrum consists of peaks at integer multiples of the phonon frequency $\omega$ relative to the ZPL frequency $\omega_{\rm ZPL}$. In the cases of off-resonant excitation, relatively broad peaks remain at these positions while additional sharp peaks at integer multiples of the phonon frequency $\omega$ relative to the laser frequency $\omega_l=\omega_{\rm ZPL}+\kappa\omega$ (dashed black lines) appear. The broad peaks constitute the background $S_{\rm is}^N(\Omega)$ from Eq.~\eqref{eq:S_bkg} due to inelastic scattering, whose position is not influenced by the detuning (colored shaded areas). The sharp peaks constitute the actual elastic photon scattering spectrum $S_{\rm es}^{N+\kappa}(\Omega)$ from Eq.~\eqref{eq:S_peak}, as they move with the detuning $\kappa\omega$ of the laser. Note that because of the initial phonon vacuum state, elastic peaks appear only below the carrier frequency of the laser.\\
	Pure dephasing was introduced in this work in a phenomenological way via a Lindblad dissipator in Eq.~\eqref{eq:D_pd}. The canonical microscopic explanation relies on processes that induce a fast spectral jitter, i.e., statistical fluctuations of the TLS frequency and thus of $\omega_{\rm ZPL}$ in the form of a stationary white-noise process, which results in phase diffusion~\cite{weiss2012quantum, bogaczewicz2023resonance}. A possible origin in our considered systems is the presence of mobile charge carriers in the vicinity of the emitter~\cite{dinu2020quantum}. This picture is helpful to understand the existence of inelastic scattering processes for $\gamma_{\rm pd}\neq 0$. Fluctuations of the frequency $\omega_{\rm ZPL}$ around some mean $\omega_{\rm ZPL}^0$ directly translate to the energy of $\ket{j_2}_X$, which is $\hbar\omega_{\rm ZPL}+\hbar j_2\omega$. This in principle allows for the absorption of a photon of frequency $\omega_{\rm ZPL}^0+\kappa\omega$, when going from the state $\ket{i}_G$ to $\ket{j_2}_X$, even for $\kappa\neq j_2-i$, since there is the possibility that $\omega_{\rm ZPL}^0+\kappa\omega=\omega_{\rm ZPL}+j_2\omega-i\omega$ due to the fluctuation of $\omega_{\rm ZPL}$. The variance of the fluctuations is directly connected to the pure dephasing rate $\gamma_{\rm pd}$~\cite{preuss2022resonant, bogaczewicz2023resonance} and these inelastic processes become more relevant for increasing $\gamma_{\rm pd}$ [compare Eq.~\eqref{eq:average_G_total} in the limit of vanishing phonon coupling $\gamma\rightarrow 0$ with Ref.~\cite{bogaczewicz2023resonance}].\\
	As discussed before, the background is in general asymmetric, but symmetric for the special case of $\text{Im}(\phi_N)=0$ with $\phi_N$ given in Eq.~\eqref{eq:phi_N}. In the limiting case of slow decay with $\gamma_{\rm xd}\ll\omega$ we have the restriction $q=n$, i.e., $j_1=j_2$ under the sum in Eq.~\eqref{eq:phi_N}. In this limit, $\phi_N$ is real and the background $S^N_{\rm is}$ is symmetric around $\Omega=\omega_{\rm ZPL}+N\omega$. Relaxing this limit a bit, the next most important contributions stem from $q=n\pm 1$, i.e., $j_1=j_2\pm 1$, which lead to a non-vanishing imaginary part of $\phi_N$ and thus to asymmetric background peaks. In physical terms, the contributions with $j_1=j_2\pm 1$ can be interpreted as interference between scattering processes into the intermediate states $\ket{j_2}_X$ and $\ket{j_2\pm 1}_X$. In the limit $\gamma_{\rm xd}\ll\omega$, these resonances have enough spectral separation to have negligible interference.
	\subsection{Full spectrum for weak dissipation and resonant excitation on ZPL or PSB}
	As already discussed in the previous two sections, the final result from Eq.~\eqref{eq:S_total_complicated} can be simplified substantially. Here, for the case of excitation on a PSB or the ZPL, i.e., $\kappa\in\mathbb{Z}$, and weak dissipation with $(\gamma_{\rm pd}+\gamma_{\rm xd})\ll\omega$, we obtain

		\begin{align}
		S(\Omega;\Gamma)&=
		\sum_{i,f} \frac{|\mathcal{E}_0|^2}{\hbar^2(\gamma_{\rm pd}+\gamma_{\rm xd})^2}|M_{i+\kappa}^fM_{i+\kappa}^{i}|^2 \rho_{i,i}^G \notag\\
		&\times \left\{ \frac{\Gamma^2}{\Gamma^2+\left[\Omega-\omega_{\rm ZPL}-(\kappa+i-f)\omega\right]^2}\label{eq:S_narrow}
		+\frac{\gamma_{\rm pd}}{\gamma_{\rm xd}}\frac{\Gamma\left(\Gamma+\frac{\gamma_{\rm pd}+\gamma_{\rm xd}}{2}\right)}{\left(\Gamma+\frac{\gamma_{\rm pd}+\gamma_{\rm xd}}{2}\right)^2\!\!\!\!+\left[\Omega-\omega_{\rm ZPL}-(\kappa+i-f)\omega\right]^2} \right\}\,,
		\end{align}
	where we have used
	\begin{equation}
	\left[\frac{\gamma_{\rm pd}+\gamma_{\rm xd}}{2}+i\omega(\kappa-k)\right]^{-1}\approx \frac{2}{\gamma_{\rm pd}+\gamma_{\rm xd}}\delta_{\kappa,k}\,.
	\end{equation}
	Each sharp peak of width $\Gamma$ (1st term in curly bracket) is accompanied by a broader Lorentzian background (2nd term in curly bracket). It might seem to be a problem, that this background diverges for $\gamma_{\rm xd}\rightarrow 0$. However, in the derivation of the RF spectrum in Sec.~\ref{sec:RF_deriv}, the stationary time-limit $(t-t_0)\gg \gamma_{\rm xd}^{-1}$ was taken, such that the limit $\gamma_{\rm xd}\rightarrow 0$ is not permissible here. The ratio of the area of the background relative to the total area of the spectrum can be inferred from Eq.~\eqref{eq:S_narrow} as 
	\begin{equation}
	\frac{\int\text{d}\Omega\, \sum_N S^N_{\rm is}(\Omega;\Gamma)}{\int\text{d}\Omega\, S(\Omega;\Gamma)}=\frac{\frac{\gamma_{\rm pd}}{\gamma_{\rm xd}}}{1+\frac{\gamma_{\rm pd}}{\gamma_{\rm xd}}}=\frac{\gamma_{\rm pd}}{\gamma_{\rm pd}+\gamma_{\rm xd}}\,.
	\end{equation}
	From this ratio, we see that the spectrum is background-dominated for $\gamma_{\rm pd}\gg\gamma_{\rm xd}$ and the background is suppressed for $\gamma_{\rm pd}\ll\gamma_{\rm xd}$.\\
	From Eq.~\eqref{eq:S_narrow} it is evident that there is a simple and direct connection between the peak weights and the initial phonon occupations, mediated by the FC factors in this important special case. This simple connection can be used to easily gain information on the initial phonon state via RF experiments. To be specific, the weight of the $k$-th peak at $\Omega=\omega_{\rm ZPL}+k\omega$ is proportional to
	\begin{equation}
	A_k=\underset{k=\kappa+i-f}{\sum_{i,f}} |M_{i+\kappa}^fM_{i+\kappa}^{i}|^2 \rho_{i,i}^G\,.\label{eq:S_amplitude_narrow}
	\end{equation}	
	\subsection{Semiclassical limit}\label{sec:classical}
	The formulas for RF spectra derived here, i.e., Eqs.~\eqref{eq:S_peak} and \eqref{eq:S_bkg} or the simplified version from Eq.~\eqref{eq:S_narrow} are fairly simple to implement, as they are basically sums of FC factors, Lorentzians and initial phonon occupations. The numerical evaluation of these sums is quick, if the initial phonon distribution is not too broad, i.e., if $\rho_{i,i}^G\approx 0$ for all except a few~$i$. This makes these formulas suitable for modeling RF spectra in the quantum limit but unsuitable in the semiclassical limit, which is usually reached for an initial coherent phonon state $\rho^G=\ket{\alpha}\bra{\alpha}$ with $|\alpha|\rightarrow\infty$, while letting the phonon coupling go to zero $|\gamma|\rightarrow 0$ and keeping $|\alpha\gamma|$ constant~\cite{hahn2022photon}. This semiclassical limit can for example describe the case of an emitter coupling to a traveling SAW~\cite{weiss2021optomechanical, wigger2021remote}. Nonetheless, we can make further analytical progress in this case, as will be demonstrated in the following.\\
	For simplicity, we perform the limit for the case of weak dissipation and resonant excitation, discussed in the previous section. That is, we aim to simplify Eq.~\eqref{eq:S_narrow}. All we need to focus on then, is how the weight of the $k$-th peak $A_k$ in Eq.~\eqref{eq:S_amplitude_narrow} behaves in the semiclassical limit. Since we consider an initial coherent state with amplitude $\alpha$, the initial phonon occupations follow the Poisson distribution~\cite{glauber1963coherent}
	\begin{equation}
	\rho_{i,i}^G=\exp(-|\alpha|^2)\frac{|\alpha|^{2i}}{i!}\,.
	\end{equation}
	As we want to understand the limit $\gamma\rightarrow 0$, we can approximate the FC factors by their leading order in $\gamma$, given in Sec.~S3 of the SI, such that
		\begin{align}
		A_k&\approx\underset{k=\kappa+i-f}{\sum_{i,f}}\frac{|\gamma|^{2|i+\kappa-f|}}{(|i+\kappa-f|!)^2}\frac{\max(i+\kappa,f)!}{\min(i+\kappa,f)!}\frac{|\gamma|^{2|i+\kappa-i|}}{(|i+\kappa-i|!)^2}\frac{\max(i+\kappa,i)!}{\min(i+\kappa,i)!}\exp(-|\alpha|^2)\frac{|\alpha|^{2i}}{i!}\notag\\
		&=\left(\frac{|\gamma|^{|k|+|\kappa|}}{|k|!|\kappa|!}\right)^2\sum_i\frac{\max(i+\kappa,i+\kappa-k)!}{\min(i+\kappa,i+\kappa-k)!}\frac{\max(i+\kappa,i)!}{\min(i+\kappa,i)!}\exp(-|\alpha|^2)\frac{|\alpha|^{2i}}{i!}\,.\label{eq:A_k_small_coupling}
		\end{align}
	Now, we need to distinguish four cases, which are $\kappa\lessgtr0$, i.e., excitation below or above the ZPL, and $k\lessgtr0$, i.e., whether we calculate the amplitude for the PSBs energetically below or above the ZPL. These cases can be extended to, e.g., $k\leq0$ and $\kappa\leq0$ to include the ZPL, without changing any of the calculations. The four cases are given in Sec.~S4 of the SI and involve different generalized hypergeometric functions~\cite{NIST:DLMF}. In the limit of $|\alpha|\rightarrow \infty$ however, all four cases have the same asymptotic behavior, resulting in
	\begin{equation}\label{eq:semiclassical}
	A_k\approx \left(\frac{D^{|k|+|\kappa|}}{|k|!|\kappa|!}\right)^2\,,\quad D=|\alpha\gamma|\,.
	\end{equation}
	In the context of SAW-driven emitters, the parameter $D$ is a measure for the strain-induced energy shift of the transition frequency $\omega_{\rm TLS}$ relative to the frequency $\omega_{\rm SAW}$ of the SAW itself~\cite{ wigger2021resonance, wigger2021remote, hahn2022photon}, which corresponds to the phonon frequency $\omega$ in our model. The weights of the peaks  thus simply follow a squared Poisson distribution with mean value $D$, which is mirrored at the ZPL ($k=0$). The $\kappa$- and $k$-dependence of the amplitude is the same, implying that optical excitation is most efficient for $|\kappa|\approx D$, i.e., at the PSB where emission is strongest. Since we are in the semiclassical limit, spontaneous emission of phonons should not be relevant. This is clearly satisfied here due to the symmetry in $\kappa$ (optical excitation) and in $k$ (light emission). This symmetry shows that phonon absorption and phonon emission are equally likely, which is only possible, if spontaneous emission of phonons is irrelevant compared to induced emission and absorption. From this reasoning we can see that any deviation from a symmetric spectrum with respect to the laser frequency indicates that the underlying phonon state is non-classical when considering the coupling with a single phonon mode. When coupling the TLS to multiple modes, asymmetric spectra might also emerge in the classical limit due to wave-mixing processes~\cite{weiss2021optomechanical}.\\
	Note, that while in general the shape of the spectrum changes with the detuning $\kappa$, in the semiclassical approximation (for resonant excitation with $\kappa\in\mathbb{Z}$) only the overall amplitude depends on $\kappa$, not the shape since we can factorize the $k$- and $\kappa$-dependence in Eq.~\eqref{eq:semiclassical}.
	\begin{figure}[t]
		\centering
		\includegraphics[width=0.5\linewidth]{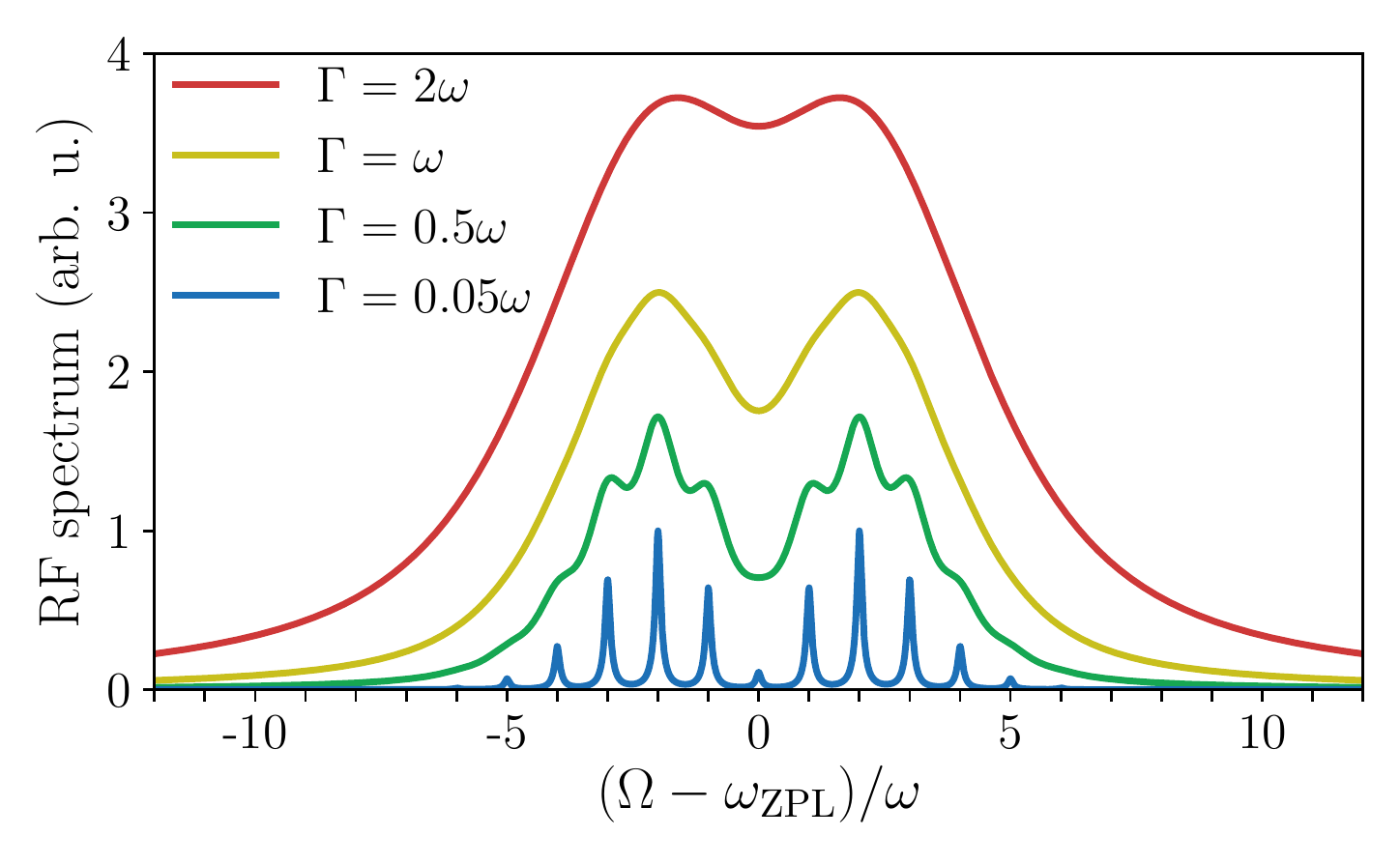}
		\caption{Examples of RF spectra in the semiclassical approximation, i.e., Eq.~\eqref{eq:S_narrow} with Eq.~\eqref{eq:S_amplitude_narrow} replaced by Eq.~\eqref{eq:semiclassical}, for different values of the spectrometer resolution $\Gamma$ as indicated in the legend. The strain-induced energy shift is $\textit{D}=\text{2.5}$, the dissiption rates are $\gamma_\text{pd}=\gamma_\text{xd}=\text{0.05}\omega$ and we consider ZPL-excitation with $\kappa=0$.}
		\label{fig:semi}
	\end{figure}\\
	Figure~\ref{fig:semi} shows examples of semiclassical RF spectra, i.e., using Eq.~\eqref{eq:S_narrow} with Eq.~\eqref{eq:S_amplitude_narrow} replaced by Eq.~\eqref{eq:semiclassical}. The strain-induced energy shift is chosen to $D=2.5$, the dissipation rates are $\gamma_{\rm pd}=\gamma_{\rm xd}=0.05\omega$ and we consider ZPL-excitation with $\kappa=0$. The graphs show RF spectra for different choices of spectrometer resolution from $\Gamma=0.05\omega$, i.e., $\Gamma\ll\omega$ (blue line), to $\Gamma=2\omega$, i.e., $\Gamma \gg \omega$ (red line). In all displayed cases we see that the maxima of the spectra are at $\Omega\approx \omega_{\rm ZPL}\pm D\omega$, in line with the previous discussion. In the case of $\Gamma\ll\omega$ (blue line), sharp peaks at integer multiples of the phonon frequency $\omega$ relative to the ZPL frequency $\omega_{\rm ZPL}$ can be seen. Their distribution is symmetric around $\omega_{\rm ZPL}$ as is expected for a single semiclassical phonon mode. Going to lower spectrometer resolution, the sharp spectral features vanish and the individual lines merge until for $\Gamma=2\omega$ (red line) the typical camelback spectrum emerges, which can also be observed in photoluminescence measurements of SAW-driven quantum dots~\cite{weiss2018interfacing, vogele2020quantum} or mechanical resonators~\cite{yeo2014strain}.\\
	Figure~\ref{fig:semi} also displays a general behavior of our spectroscopy model. The total intensity of the scattered light, i.e., the area under the curves in Fig.~\ref{fig:semi}, increases with increasing $\Gamma$, i.e., decreasing resolution of the spectral filter. This can already be seen from the time-integrated spectrum in Eq.~\eqref{eq:S_RF}. The integral over $\Omega$ of this spectrum would enforce $\tau=0$, leading to
	\begin{equation}\label{eq:freq_int_S}
	\int\dd\Omega\, S(\Omega;\Gamma)\sim \Gamma\overline{G}(0)\,.
	\end{equation}
	The time-averaged correlation function at $\tau=0$ is here given by
	\begin{equation}
	\overline{G}(0)=\lim\limits_{T\rightarrow\infty}\frac{1}{T}\int\limits_{t_0}^{t_0+T}\text{d}t\,\expval{X^{\dagger}X}(t)
	\end{equation}
	and is thus determined by the stationary occupation of the TLS emerging from a balance between optical driving and decay, as discussed in the context of Eq.~\eqref{eq:rho_X}. The total intensity thus scales with the stationary occupation of the emitter and the resolution of the spectral filter. This is physically sound, since perfect resolution $\Gamma\rightarrow 0$ would require a resonator with perfect mirrors in the case of our prototypical spectrometer model, such that the incoming light would need infinitely long to get to the photon detector~\cite{eberly1977time}.\\
	It might seem to be a problem, that Eq.~\eqref{eq:freq_int_S} diverges in the limit $\Gamma\rightarrow \infty$. Note however, that in this limit the time-integrated spectrum from Eq.~\eqref{eq:S_RF} is given by
	\begin{equation}
	\lim\limits_{\Gamma\rightarrow\infty} S(\Omega;\Gamma)=\overline{G}(0)\,,
	\end{equation}
	i.e., the spectrum becomes independent of the setting frequency $\Omega$ and is just determined by the stationary occupation of the TLS. If we then additionally integrated over the setting frequency $\Omega$, we would of course obtain a result which diverges  simply due to the independence of the spectrum from $\Omega$, as can be seen in Eq.~\eqref{eq:freq_int_S} for $\Gamma\rightarrow\infty$. 
	\section{Reading phonon number statistics from RF spectra}\label{sec:reading}
	After thoroughly characterizing our model, we proceed to the central part of this work. In the following we demonstrate, how to read the unknown initial phonon number statistics from the RF spectrum in the case of optical excitation on a PSB or the ZPL with $\kappa\in \mathbb{Z}$. We start by using the simple formula in Eq.~\eqref{eq:S_amplitude_narrow}, which will make the underlying physics obvious. As no experiments are available yet, we need to simulate the RF spectra, that shall be fitted. To do so, we first generate a RF spectrum by performing the full simulation via Eq.~\eqref{eq:S_total_complicated} for a given initial phonon state, with phonon occupations $\rho_{i,i}^G$. We then fit this spectrum with Eq.~\eqref{eq:S_narrow} using a non-linear least squares algorithm~\cite{more2006levenberg}. We assume that the phonon coupling $\gamma=g/\omega$, and thus the FC factors, are already well-known, e.g., from absorption or emission spectra. For simplicity, we also assume that the dephasing and decay rates, as well as the spectrometer resolution are known. Consequently, the only variables in Eq.~\eqref{eq:S_narrow}, up to an overall prefactor, are the phonon occupations $\rho^G_{i,i}$. As a measure for the deviation of the fitted phonon occupations $\rho^G_{i,i;{\rm fit}}$ from the initial $\rho_{i,i}^G$ entering Eq.~\eqref{eq:S_total_complicated}, we use
	\begin{equation}\label{eq:delta_fit}
	\Delta_{\rm read}=\frac{1}{2}\sum_i \left|\rho^G_{i,i}-\rho^G_{i,i;{\rm fit}}\right|\,,
	\end{equation}
	whose bounds are given by
	\begin{align}
	0\leq \Delta_{\rm read} &\leq \frac{1}{2}\sum_i \left|\rho^G_{i,i}\right|+\frac{1}{2}\sum_i\left|\rho^G_{i,i;{\rm fit}}\right| \notag\\
	&=\frac{1}{2}\text{Tr}(\rho^G)+\frac{1}{2}\text{Tr}(\rho^G_{\rm fit})=1,
	\end{align}
	with the upper bound in the first line being the value that is obtained for the case that $\rho_{i,i}^G$ is exactly zero, when $\rho_{i,i;{\rm fit}}^G$ is not and vice versa. Note, that the definition of $\Delta_{\rm read}$ coincides with the trace distance between the density matrices $\rho^G$ and $\rho^G_{\rm fit}$, if both are diagonal~\cite{nielsen2002quantum}. This quantity measures how well one can extract the information on the phonon occupations $\rho^G_{i,i}$ from the fitting procedure. As starting parameters for the fit we choose the unphysical values $\rho_{i,i;{\rm start}}^{G}=0$. After the fitting procedure the parameters are normalized to $\text{Tr}(\rho^G_{\rm fit})=1$. Note, that $\Delta_{\rm read}$ has nothing to do with the standard deviation of the fitting procedure and is a quantity derived from the fitted parameters to quantify the information obtained from the fit, i.e., the readout of the phonon occupations. In the following we will call it the readout error.\\
	Before we start with the fit of the spectra using Eq.~\eqref{eq:S_narrow}, we need to consider that any realistic RF spectrum will contain a finite number of recognizable and distinguishable peaks. Any measured spectrum will also have an unavoidable background in addition to the background from the actual photon scattering, discussed previously in Sec.~\ref{sec:bkg}. The phononic Hilbert space however, is infinite-dimensional and thus there are in principle an infinite number of free parameters $\rho_{i,i}^G$ in Eq.~\eqref{eq:S_narrow}. To tackle this issue, we consider the energetically highest, visible peak at frequency $\Omega=\omega_{\rm ZPL}+\kappa\omega+N_{\rm max}\omega$. According to Eq.~\eqref{eq:S_peak} and the discussion in Sec.~\ref{sec:RF_qual_deriv}, this peak stems from a process of scattering from an initial state $\ket{i}_G=\ket{N_{\rm max}}_G$ into a final state $\ket{f}_G=\ket{0}_G$, i.e., absorption of $N_{\rm max}$ phonons during the photon scattering process. Also processes like $\ket{N_{\rm max}+1}_G\rightarrow\ket{1}_G$ can lead to such a peak. However, then we would in general also expect a peak at $\Omega=\omega_{\rm ZPL}+\kappa\omega+(N_{\rm max}+1)\omega$, in contradiction to the initial assumption. However, in special cases, e.g., when the transition  $\ket{N_{\rm max}+1}_G\rightarrow\ket{0}_G$ is forbidden due to a vanishing FC factor, the argument from above breaks down. We will come back to this issue later.\\
	For the moment we can thus assume that the energetically highest peak at  $\Omega=\omega_{\rm ZPL}+\kappa\omega+N_{\rm max}\omega$ dictates the maximum number of phonons to consider for the fit of the initial phonon density matrix $\rho^G_{\rm fit}$. We have then in total $N_{\rm max}$ phonon absorption peaks at frequencies $\Omega=\omega_{\rm ZPL}+\kappa\omega+N\omega$ with $N=1, ..., N_{\rm max}$ and an arbitrary number of phonon emission peaks below $\Omega=\omega_{\rm ZPL}+\kappa\omega$, as well as the ZPL at $\Omega=\omega_{\rm ZPL}+\kappa\omega$. The spectrum thus contains enough information to determine $\rho^G$ in the truncated Hilbert space spanned by the phonon states $\{\ket{0}$, \dots, $\ket{N_{\rm max}}\}$.\\	
	We consider the example of a superposition of the phononic vacuum $\ket{0}$ and the single phonon state $\ket{1}$. To be specific, we consider the initial phonon density matrix 
	\begin{equation}\label{eq:rho_G_10}
	\rho^G=\frac{1}{2}\left(\ket{0}\bra{0}+\ket{1}\bra{0}+\ket{0}\bra{1}+\ket{1}\bra{1}\right)\,.
	\end{equation} 
	Although only the phonon occupations enter the RF spectra in Eqs.~\eqref{eq:S_total_complicated} and ~\eqref{eq:S_narrow}, we consider here a pure state including coherences for the sake of presentation clarity. We will make use of Wigner functions of the excited state subspace $W_X$ in the following discussion~\cite{raimond2006exploring}. This Wigner function is defined as the two-dimensional Fourier transform of the symmetrically ordered phonon characteristic function and can be expressed as [Eq.~(S35) in Sec.~S3 of the SI with $\rho\rightarrow\bra{X}\rho(t)\ket{X}$]
	\begin{align}
	W_X(\alpha)&=\sum_{m,n}\bra{m}\bra{X}\rho\ket{X}\ket{n} (-1)^m \frac{2}{\pi}\bra{n} D(2\alpha)\ket{m}\,.
	\end{align}
	Here, $D$ is the unitary displacement operator~\cite{glauber1963coherent} and the phase space is spanned by $(\text{Re}\,\alpha,\text{Im}\,\alpha)$. This expression can also be written in terms of the eigenstates $\ket{m}_X$ of $H_0$ from Eq.~\eqref{eq:H0_spec_x}
	\begin{align}\label{eq:W_X}
	W_X(\alpha)&=\sum_{m,n}\braind{X}{m}\rho\ket{n}_X (-1)^m \frac{2}{\pi}\bra{n} D(2\alpha+2\gamma)\ket{m}\,,
	\end{align}
	where we used that an active unitary transformation of the density matrix with a displacement operator $\rho\rightarrow D(\gamma)\rho D(-\gamma)$ simply leads to a displacement of the corresponding Wigner function by $\gamma$, i.e., $\alpha\rightarrow\alpha-\gamma$ in its argument, as can be inferred easily from Sec.~S3 of the SI. Here, we can insert the matrix elements from Eq.~\eqref{eq:rho_X} to obtain the Wigner function of the excited state subspace. The visual identification of the phonon state associated with the excited state of the TLS via Wigner functions is much more straightforward when coherences are included. These, however, do not have any impact on the presented RF spectra or the readout of the number statistics.
	\subsection{ZPL excitation and weak dissipation}
\begin{figure}[h]
		\centering
		\includegraphics[width=0.48\linewidth]{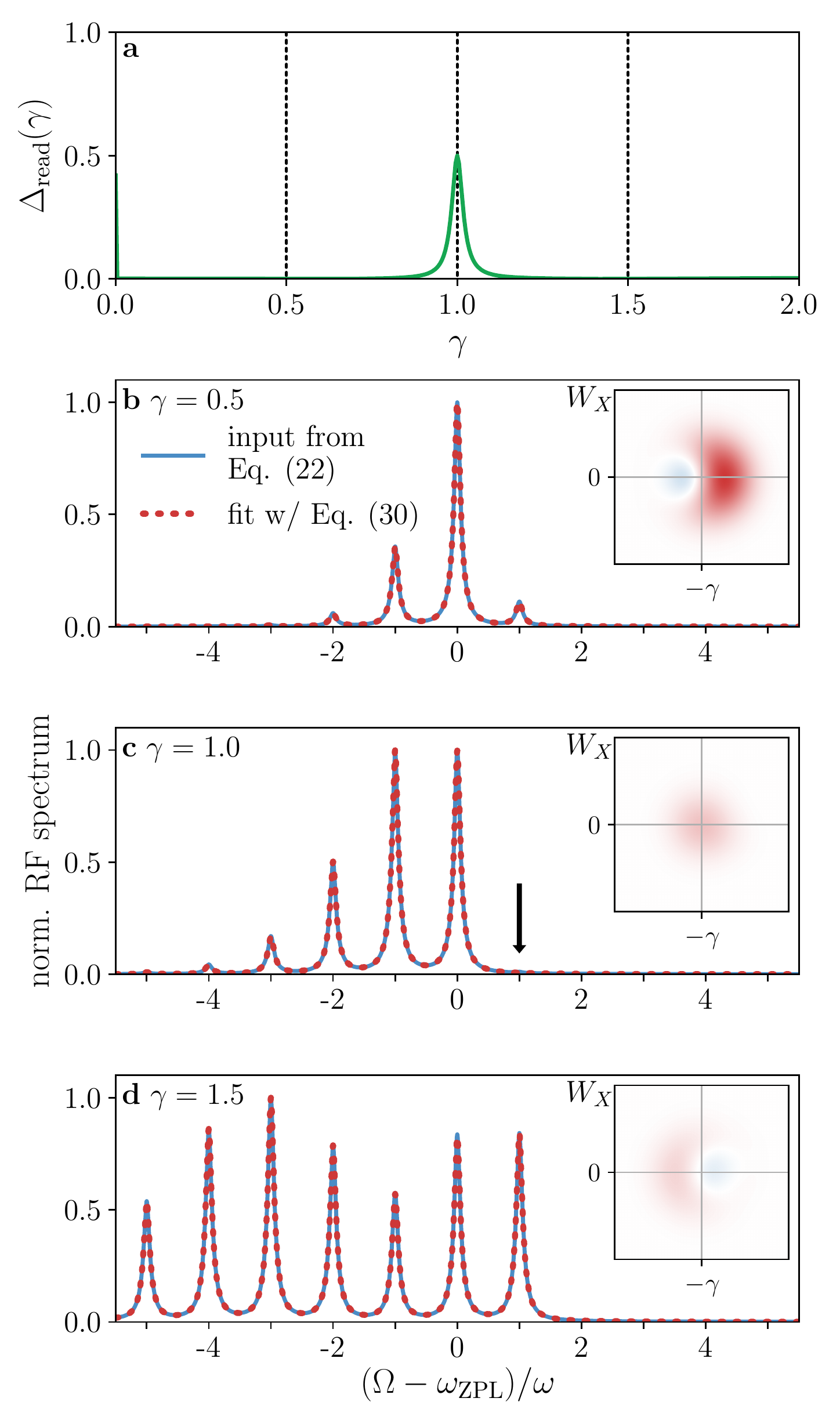}
		\caption{Fit of the full RF spectrum from Eq.~\eqref{eq:S_total_complicated} using the approximation from Eq.~\eqref{eq:S_narrow} for a symmetric superposition of initial phonon states $\ket{\text{0}}$ and $\ket{\text{1}}$. (a) Readout error $\Delta_\text{read}$, determined with Eq.~\eqref{eq:delta_fit} depending on the phonon coupling $\gamma$. (b)-(d) Full spectra (blue, solid) and fits (red, dashed) for special values of $\gamma$ as marked by vertical lines in (a). Insets: Wigner functions $\textit{W}_\textit{X}$ [Eq.~\eqref{eq:W_X}] of the excited state density matrix from Eq.~\eqref{eq:rho_X} at full phonon periods $\omega(t-t_0)/(\text{2}\pi)\in\mathbb{N}$, plotted on a common color scale. Blue colors correspond to negative and red colors to positive values of the Wigner function. The parameters of the model are $\kappa=0$ (ZPL-excitation) and $\gamma_\text{pd}=\gamma_\text{xd}=\Gamma=\text{0.05}\omega$.}
		\label{fig:tracedistance10}
	\end{figure}

	We consider first excitation on the ZPL, i.e., $\kappa=0$, and assume that we have identified a peak at $\Omega=\omega_{\rm ZPL}+1\omega$, such that the maximum number of phonons can be approximated as $N_{\rm max}=1$. Fig.~\ref{fig:tracedistance10}(a) shows the readout error $\Delta_{\rm read}$ as a function of the phonon coupling $\gamma$ for fixed $\gamma_{\rm pd}=\gamma_{\rm xd}=\Gamma=0.05\omega$. In almost the entire considered range of phonon couplings we find $\Delta_{\rm read}\approx 0$, meaning that we are able to gain accurate information on the phonon occupations, except for (i) $\gamma\approx 0$ and (ii) $\gamma\approx 1$, where we clearly have $\Delta_{\rm read}> 0$.\\
	We focus first on the case (i) of $\gamma\rightarrow 0$. In this regime, the coupling of the phonons to the emitter is so weak, that no information on the initial phonon state can be obtained by scattering light off the emitter. In the extreme case of $\gamma=0$, the phonons are completely decoupled from the TLS, such that the light scattering cannot influence the phonons and cannot be influenced by them. Thus, any phonon density matrix will lead to the same RF spectrum (which contains only the ZPL in this limit). Therefore, to be able to gain reliable information on the initial phonon state via light scattering, we need to have a sufficiently strong coupling $\gamma$.\\
	Case (ii) is more subtle. To understand it better, Figs.~\ref{fig:tracedistance10}(b)-(d) show the full input RF spectra calculated with Eq.~\eqref{eq:S_total_complicated} (solid blue lines) together with the fits using Eq.~\eqref{eq:S_narrow} (dashed red lines) for selected values of $\gamma$ as indicated by the vertical dashed lines in (a). In addition to each RF spectrum, the Wigner function $W_X$ [Eq.~\eqref{eq:W_X}] of the excited state's phonon density matrix from Eq.~\eqref{eq:rho_X} is shown in the insets with blue colors indicating negative values and red colors indicating positive values. For its calculation we use the full initial state from Eq.~\eqref{eq:rho_G_10} in Eq.~\eqref{eq:rho_X}. The Wigner function thus shows which part of the initial phonon state [Eq.~\eqref{eq:rho_G_10}] can be transferred to the excited state subspace of the TLS [Eq.~\eqref{eq:rho_X}] by photon absorption and can therefore contribute to the photon scattering process. All Wigner functions are plotted on a common color scale. In all three cases, the fitted spectra and the full calculations match excellently. This is due to the fact that we are in the regime of $\gamma_{\rm pd}+\gamma_{\rm xd}\ll\omega$, where the approximation from Eq.~\eqref{eq:S_narrow} is applicable.\\
	The value of $\gamma$ in Figs.~\ref{fig:tracedistance10}(b)-(d) increases from top to bottom. At the top in (b) for $\gamma=0.5$ we see a typical RF spectrum with a strong ZPL at $\Omega=\omega_{\rm ZPL}$. In addition, there are a single phonon absorption PSB at $\Omega=\omega_{\rm ZPL}+\omega$ and two well-visible phonon emission PSBs below the ZPL. The general behavior on the emission side when going from $\gamma=0.5$ in (b) to $\gamma=1.5$ in (d) is that the number of discernible peaks and their overall amplitude increases. This is due to the fact that with stronger phonon coupling, multi-phonon emission becomes more likely. This behavior is not reproduced on the absorption side. First of all, since the maximum number of phonons is one, no additional PSBs above $\Omega=\omega_{\rm ZPL}+\omega$ can appear. Furthermore, when comparing the case of $\gamma=0.5$ in (b) with $\gamma=1.5$ in (d) we see that the amplitude of the absorption PSB has increased dramatically. Contrary to what might be expected however, the amplitude does not increase monotonically, as can be seen in (c) for $\gamma=1$, where the absorption PSB, whose position is marked by the black arrow, vanishes completely.\\
	To get a deeper understanding of the peculiar behavior for $\gamma=1$ in Fig.~\ref{fig:tracedistance10}(c), we first consider the impact of the phonon coupling $\gamma$ on the excited state density matrix from Eq.~\eqref{eq:rho_X}. In the limit $(\gamma_{\rm pd}+\gamma_{\rm xd})\ll\omega$, which holds here, and for $\kappa=0$, the matrix elements are given by
	\begin{equation}\label{eq:elements_resonant}
	\braind{X}{m}\rho\ket{n}_X\sim M_m^m M_n^{n*}\rho_{m,n}^G\,.
	\end{equation}
	Thus, the information on the initial phonon state is transferred from $\ket{G}$ to $\ket{X}$ via the FC factors $M_m^m=\bra{m}B_+\ket{m}=\braind{X}{m}X^{\dagger}\ket{m}_G$, describing the optical transition $\ket{m}_G\rightarrow\ket{m}_X$. As derived in Sec.~S3 in the SI, these special FC factors are given by
	\begin{equation}
	M_m^m=\exp(-\frac{|\gamma|^2}{2})L_m(|\gamma|^2)
	\end{equation}
	with $L_m$ being the $m$-th Laguerre polynomial~\cite{NIST:DLMF}. Due to the Gaussian prefactor the excited state density matrix elements are damped with increasing $\gamma$ for a sufficiently large $\gamma$, where the Gaussian dominates the Laguerre polynomial. Thus, the excitation of the emitter becomes inefficient for $\gamma\rightarrow\infty$. This can already be seen in Fig.~\ref{fig:tracedistance10}(b)-(d) by the color intensity of the Wigner functions decreasing with increasing $\gamma$. The appearance of the Laguerre polynomials explains the anomalous behavior for $\gamma=1$. For the initial state $\rho^G$ considered here, the only non-vanishing density matrix elements are $\rho_{0,0}^G=\rho_{1,1}^G=\rho_{0,1}^G=\rho_{1,0}^G=0.5$. Thus, according to Eq.~\eqref{eq:elements_resonant}, the only relevant FC factors are $M_0^0\sim L_0(|\gamma|^2)$ and $M_1^1\sim L_1(|\gamma|^2)$. In general, the $m$-th Laguerre polynomial $L_m$ is an $m$-th order polynomial and has $m$ roots. For the special cases relevant here, we have $L_0(|\gamma|^2)=1$ and $L_1(|\gamma|^2)=1-|\gamma|^2$. Thus, in the subspace spanned by $\ket{0}_X$ and $\ket{1}_X$ we have 
	\begin{equation}
	\bra{X}\rho\ket{X}\sim \left(\begin{matrix}1&1-|\gamma|^2\\1-|\gamma|^2&(1-|\gamma|^2)^2\end{matrix}\right)\,.
	\end{equation}
	For $\gamma\rightarrow 0$ this coincides with the initial density matrix $\rho^G$ in the basis $\qty{\ket{0},\ket{1}}$, yielding the typical Wigner function for a superposition of $\ket{0}_X$ and $\ket{1}_X$ in Fig.~\ref{fig:tracedistance10}(b). For $|\gamma|>1$, the sign of the off-diagonal matrix elements inverts, i.e., the phase between $\ket{0}_X$ and $\ket{1}_X$ is then opposite to the case $|\gamma|<1$. This leads to the mirroring of the Wigner function in (d) with respect to (b). For the special case $|\gamma|=1$, we obtain
	\begin{align}
	\bra{X}\rho\ket{X}(\gamma=1)\sim B_-\ket{0}\bra{0}B_+ = \begin{pmatrix} 1 & 0 \\ 0 & 0 \end{pmatrix}\,,
	\end{align}
	i.e., a displaced phonon vacuum, as displayed by the Wigner function in Fig.~\ref{fig:tracedistance10}(c). We see that this value for $\gamma$ is exactly the root of the polynomial $L_1$, implying $M_1^1=0$. Physically, this means that the optical transition $\ket{1}_G\rightarrow\ket{1}_X$ is forbidden in the leading order with respect to the driving field such that no information on the presence of the state $\ket{1}_G$ in the initial phonon state $\rho^G$ is transferred to the excited state here.\\
	This vanishing of the FC factor $M_1^1$ for $\gamma=1$, i.e., the presence of a forbidden optical transition, of course has implications for the shape of the RF spectrum. The weights of the peaks from Eq.~\eqref{eq:S_amplitude_narrow} can be written as 
	\begin{equation}
	A_k\sim |M_0^{-k}|^2\rho^G_{0,0}
	\end{equation}
	for the parameters from Fig.~\ref{fig:tracedistance10}(c). Since $M_0^{-k}=0$ for $k>0$, there are no absorption PSBs. No information on the initial occupation of the single phonon state $\ket{1}$ enters here. Thus, when fitting the spectrum, the occupation $\rho^G_{1,1}$ is an undetermined degree of freedom and remains at its initial value of $\rho^G_{1,1;\mathrm{start}}=0$. In principle we then cannot give a good estimate for the initial phonon state $\rho^G$, for which $\rho^G_{1,1}\neq 0$. Consequently, this leads to the peak in Fig.~\ref{fig:tracedistance10}(a) at $\gamma=1$, which we call a {\it  blind} $\gamma$-value. Note, that the fit employing Eq.~\eqref{eq:S_narrow} and the full simulation using Eq.~\eqref{eq:S_total_complicated} agree excellently in Fig.~\ref{fig:tracedistance10}(c). Nonetheless, we simply cannot obtain any information on the initial occupation of the single phonon state, leading to a large $\Delta_{\rm read}$. The standard deviation for $\Delta_{\rm read}$, as obtained from the fitting procedure, is actually negligible everywhere except around $\gamma=1$, where it diverges due to the presence of a free parameter.\\
	At this point we have to come back to our initial assumption that the energetically highest peak determines the size of the reduced Hilbert space for the phonons, when performing the fit. We have just learned that in special cases, where relevant optical transitions are forbidden (the FC factor vanishes), this assumption clearly breaks down, as can be seen in Fig.~\ref{fig:tracedistance10}(c). In these cases however, the determination of the initial phonon occupation is not possible anyways, at least when using the approximation from Eq.~\eqref{eq:S_narrow}. Thus, one should try to avoid situations, where relevant FC factors vanish. As we will see in Sec.~\ref{sec:noise}, this is still true when employing the full formula from Eq.~\eqref{eq:S_total_complicated} for the fit in more realistic situations with an additional noisy background in the spectrum.
	\subsection{Detuned excitation as a check for consistency}\label{sec:detuning}
\begin{figure}[h]
		\centering
		\includegraphics[width=0.48\linewidth]{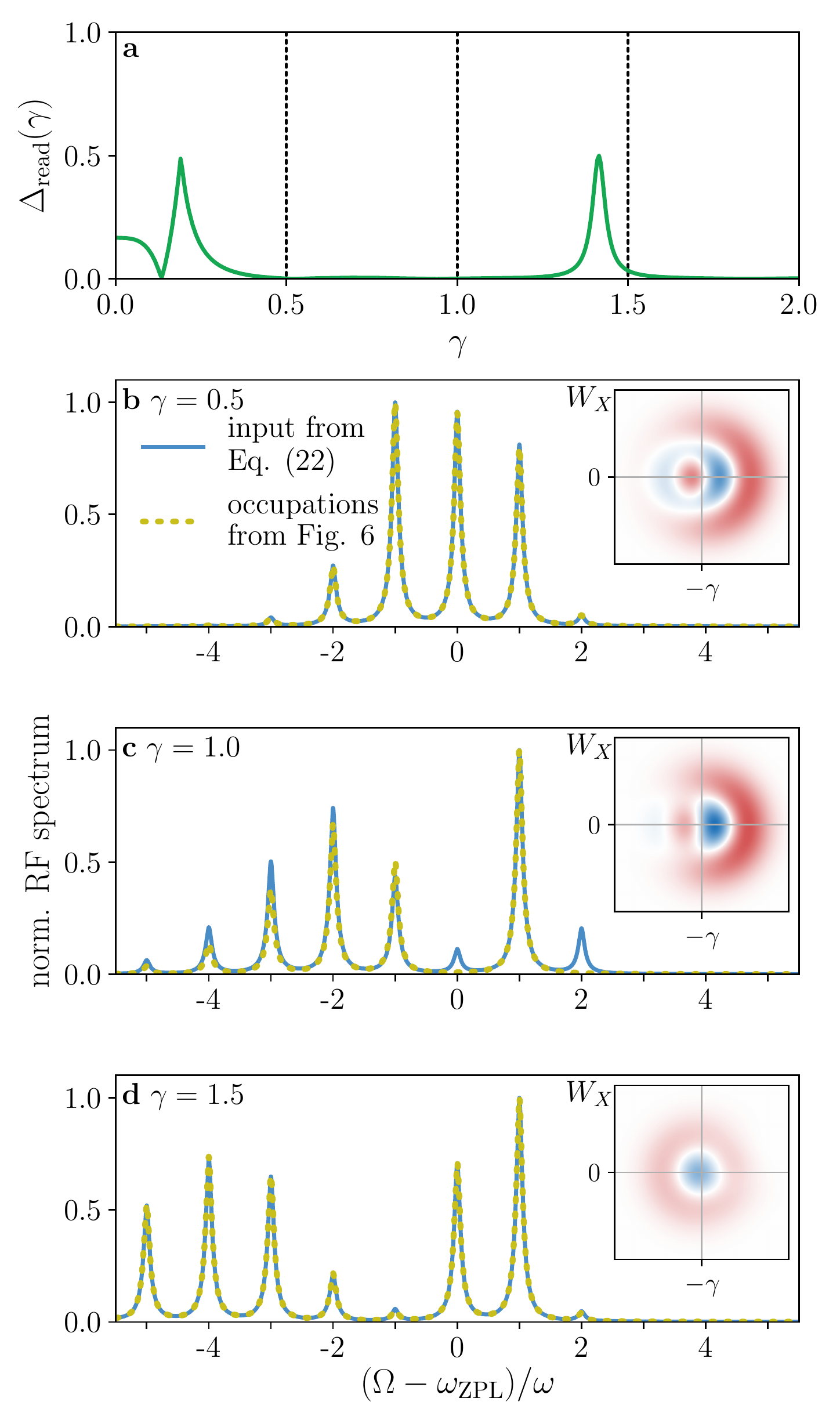}
		\caption{Detuned excitation for symmetric superposition of initial phonon states $\ket{\text{0}}$ and $\ket{\text{1}}$. (a) Readout errors $\Delta_\text{read}$, determined with Eq.~\eqref{eq:delta_fit} depending on the phonon coupling $\gamma$ for a fit of the full RF spectrum from Eq.~\eqref{eq:S_total_complicated} using the approximation from Eq.~\eqref{eq:S_narrow}.  (b)-(d) Full spectra (blue, solid) for special values of $\gamma$ as marked by vertical lines in (a). Approximative spectra employing Eq.~\eqref{eq:S_narrow} using the predicted occupations from Fig.~\ref{fig:tracedistance10} (yellow, dashed). Insets: Wigner functions $\textit{W}_\textit{X}$ [Eq.~\eqref{eq:W_X}] of the excited state density matrix from Eq.~\eqref{eq:rho_X} at full phonon periods $\omega(t-t_0)/(\text{2}\pi)\in\mathbb{N}$, plotted on a common color scale. Blue colors correspond to negative and red colors to positive values of the Wigner function. The parameters of the model are $\kappa=1$ (PSB-excitation) and $\gamma_\text{pd}=\gamma_\text{xd}=\Gamma=\text{0.05}\omega$.}
		\label{fig:tracedistance10_det1}
	\end{figure}

	We now turn to detuned excitation on the first PSB with $\kappa=1$. We again perform a full simulation using Eq.~\eqref{eq:S_total_complicated} for different phonon couplings $\gamma$ and fit the emerging spectra using Eq.~\eqref{eq:S_narrow}. Apart from the different detuning, we choose the same parameters as in Fig.~\ref{fig:tracedistance10}. The resulting readout errors $\Delta_{\rm read}$ are shown in Fig.~\ref{fig:tracedistance10_det1} (a). Again, in almost the entire considered range of phonon couplings we find $\Delta_{\rm read}\approx 0$, meaning that we are able to gain information on the phonon occupations, except for (i) $\gamma< 0.5$ and (ii) $\gamma\approx 1.4$, where we clearly have $\Delta_{\rm read}> 0$.\\
	Case (i) can be understood analog to Fig.~\ref{fig:tracedistance10} (a), where the readout became challenging for $\gamma\rightarrow 0$. However, here for PSB-excitation with $\kappa=1$ the range of large $\Delta_{\rm read}$ at $\gamma\rightarrow 0$ is much broader compared to the case of ZPL-excitation with $\kappa=0$ in Fig.~\ref{fig:tracedistance10}. To understand this difference, it is instructive to compare the strongest peak in the RF spectra with the energetically highest peak, employing Eq.~\eqref{eq:S_amplitude_narrow} and the low-$\gamma$ limit from Eq.~(S42) in the SI. The strongest peak in this limit is the ZPL with the weight 
	\begin{equation}
	A_0\sim\mathcal{O}(|\gamma|^{2|\kappa|})\,.
	\end{equation}
	The energetically highest peak, which stems from the scattering process $\ket{i=N_{\rm max}}_G\rightarrow\ket{f=0}_G$ has the weight 
	\begin{equation}
	A_{N_{\rm max}}\sim\mathcal{O}(|\gamma|^{2N_{\rm max}+4|\kappa|})\,.
	\end{equation}
	As discussed before, it contains information on the initial occupation of the energetically highest phonon state $\ket{N_{\rm max}}_G$. Thus, if it cannot be distinguished well from the rest of the spectrum, the determination of $\rho_{N_{\rm max},N_{\rm max}}^G$ becomes challenging. The ratio of the weight of the energetically highest peak with the weight of the ZPL depends on $\kappa$ and scales as
	\begin{equation}
	\frac{A_{N_{\rm max}}}{A_0}\sim\mathcal{O}(|\gamma|^{2|\kappa|})\,.
	\end{equation}
	Thus, for increasing absolute detunings $|\kappa|\omega$, we expect the determination of the initial phonon occupations to be more challenging at small values of $\gamma$ because the information on the highest occupied phonon states is getting lost from the RF spectrum, which is in line with the results in Fig.~\ref{fig:tracedistance10_det1}(a).\\
	Turning to case (ii), we see that the peak in Fig.~\ref{fig:tracedistance10_det1}(a) is at a different {\it blind} $\gamma$-value compared to Fig.~\ref{fig:tracedistance10}(a). Since the laser has a different detuning, i.e., $\kappa=1$, the resonant processes are now $\ket{i}_G\rightarrow \ket{i+1}_X$ instead of $\ket{i}_G\rightarrow\ket{i}_X$. Thus, the information on the initial occupation of $\ket{i}_G$ is transferred to the excited state, even if $\ket{i}_G\rightarrow\ket{i}_X$ is forbidden, i.e., $M_i^i=0$. Instead, the critical FC factors are here $M_i^{i+1}$ with $i=0,1$, as can also be seen from Eq.~\eqref{eq:S_amplitude_narrow} for $k>0$ (absorption PSBs). Only one of those FC factors, namely $M_1^2$, has a single root at $\gamma=\sqrt{2}$. This is precisely where the peak in Fig.~\ref{fig:tracedistance10_det1}(a) appears.\\
	Since the relevant resonant transitions are in general $\ket{i}_G\rightarrow\ket{i+\kappa}_X$, the relevant FC factors $M_{i+\kappa}^{i}$ are not only influenced by the laser detuning $\kappa\omega$, but also by the distribution of initial phonon occupations $\rho^G_{i,i}$, as can be seen from Eq.~\eqref{eq:S_amplitude_narrow}. Generally, the relevant FC factors behave as
	\begin{equation}\label{key}
	M_{i+\kappa}^i\sim L_{\min(i+\kappa,i)}^{(|\kappa|)}(|\gamma|^2)\,,
	\end{equation}
	as is derived in Sec.~S3 in the SI, where $L_{\min(i+\kappa,i)}^{(|\kappa|)}$ is a generalized Laguerre-polynomial~\cite{NIST:DLMF}. This is a polynomial of degree $\min(i+\kappa,i)$ with $\min(i+\kappa,i)$ roots. Thus, the number of roots, i.e., the number of {\it blind} $\gamma$-values, increases when phonon states with higher occupation $i$ are involved.\\
	Next, we illustrate the impact of a large value of $\Delta_{\rm read}$. To this aim, Figs.~\ref{fig:tracedistance10_det1} (b)-(d) show full spectra (blue, solid) calculated using Eq.~\eqref{eq:S_total_complicated} for special values of $\gamma$ as marked by vertical dashed lines in (a). As in Fig.~\ref{fig:tracedistance10}, the insets show the corresponding excited state Wigner functions $W_X$. In contrast to Fig.~\ref{fig:tracedistance10} however, the dashed yellow lines in (b)-(d) do not show the fit of the simulated spectra. Here they correspond to approximative spectra using Eq.~\eqref{eq:S_narrow} that one would predict based on the occupations that we obtained from the respective fits in Fig.~\ref{fig:tracedistance10}.\\
	Starting at the top in (b) with $\gamma=0.5$, we see that the prediction based on Fig.~\ref{fig:tracedistance10} and the full simulation agree excellently. This result is of course expected, since the readout error $\Delta_{\rm read}$ was zero at $\gamma=0.5$ in Fig.~\ref{fig:tracedistance10}. Also here, at a detuning of $\kappa=1$, the readout works fine, as can be seen in (a). The corresponding Wigner function shows a typical superposition of $\ket{1}_X$ and $\ket{2}_X$, implying that the information on the occupations of the initial states $\ket{i}_G$ is transferred to the excited state subspace via $\ket{i}_G\rightarrow \ket{i+\kappa}_X$.\\
	In (c) at $\gamma=1$ there is a qualitative difference between the full simulation and the prediction based on Fig.~\ref{fig:tracedistance10}. As discussed previously, for ZPL excitation at $\gamma=1$, a full readout of the initial phonon state considered here is not possible, since the transition $\ket{1}_G\rightarrow\ket{1}_X$ is forbidden for weak optical driving. Thus, the fit in Fig.~\ref{fig:tracedistance10} yields a density matrix of the form $\rho_{{\rm fit}}^G=\ket{0}\bra{0}$. When calculating RF spectra based on this density matrix for a detuning of $\kappa=1$, the energetically highest peak is at the corresponding laser frequency $\Omega=\omega_{\rm ZPL}+1\omega$. However, for the correct initial superposition of $\ket{0}$ and $\ket{1}$, the energetically highest peak is of course at $\Omega=\omega_{\rm ZPL}+2\omega$ due to phonon absorption from the initial $\ket{1}$-contribution. This leads to considerable qualitative deviations between the solid blue and the dashed yellow line in (c). Therefore, trying to read out the initial phonon number statistics for detunings $\kappa$ and couplings $\gamma$, for which relevant optical transitions are forbidden, leads to inconsistencies between predictions based on the readout and measurements of RF spectra at different detunings. If we reverse the order of readout and prediction here, i.e., if we first performed the readout at $\kappa=1$ to predict the spectrum at $\kappa=0$, such an inconsistency would not emerge for the coupling $\gamma=1$ since the corresponding error $\Delta_{\rm read}$ vanishes in Fig.~\ref{fig:tracedistance10_det1} (a). The corresonding Wigner function in the excited state $W_X$ also shows a typical superposition of $\ket{1}_X$ and $\ket{2}_X$, consistent with the fact that the information on the initial phonon state is transferred to the excited state subspace for PSB-excitation with $\kappa=1$ and a phonon coupling of $\gamma=1$, while this was not true for ZPL-excitation in Fig.~\ref{fig:tracedistance10} (c).\\
	Finally, in Fig.~\ref{fig:tracedistance10_det1} (d) for a coupling of $\gamma=1.5$, the full simulation and the prediction due to Fig.~\ref{fig:tracedistance10} coincide again since the error $\Delta_{\rm read}$ was zero for this value of $\gamma$ in Fig.~\ref{fig:tracedistance10} (a). Since the \textit{blind} $\gamma$-value for a detuning of $\kappa=1$ is at $\gamma=\sqrt{2}$, i.e., close to $\gamma=1.5$, the readout would not work perfectly here, as can be seen in Fig.~\ref{fig:tracedistance10_det1} (a) due to the nonzero $\Delta_{\rm read}$. This is underlined by the fact that the Wigner function in (d) almost looks like a pure $\ket{1}_X$ state, apart from a slight deformation. At the \textit{blind} spot of $\gamma=\sqrt{2}$ this deformation would vanish, leaving only a contribution due to the transition $\ket{0}_G\rightarrow\ket{1}_X$, while the transition $\ket{1}_G\rightarrow\ket{2}_X$ is forbidden, i.e., $M_1^2(\gamma=\sqrt{2})=0$. Analog to the discussion in (c), we would obtain qualitative differences when reversing the order of readout and prediction between Fig.~\ref{fig:tracedistance10} (d) and Fig.~\ref{fig:tracedistance10_det1} (d) since the readout does not work perfectly for detuned excitation with $\kappa=1$ at a coupling of $\gamma=1.5$.\\
	To conclude this section, we have seen that a different detuning leads to different \textit{blind} $\gamma$-values, which can be used to work around these critical values. Furthermore, performing measurements at different detunings can be used as a check for consistency of the phonon number statistics readout.
	\subsection{Impact of noise and dissipation}\label{sec:noise}
	\begin{figure}[h]
		\centering
		\includegraphics[width=0.5\linewidth]{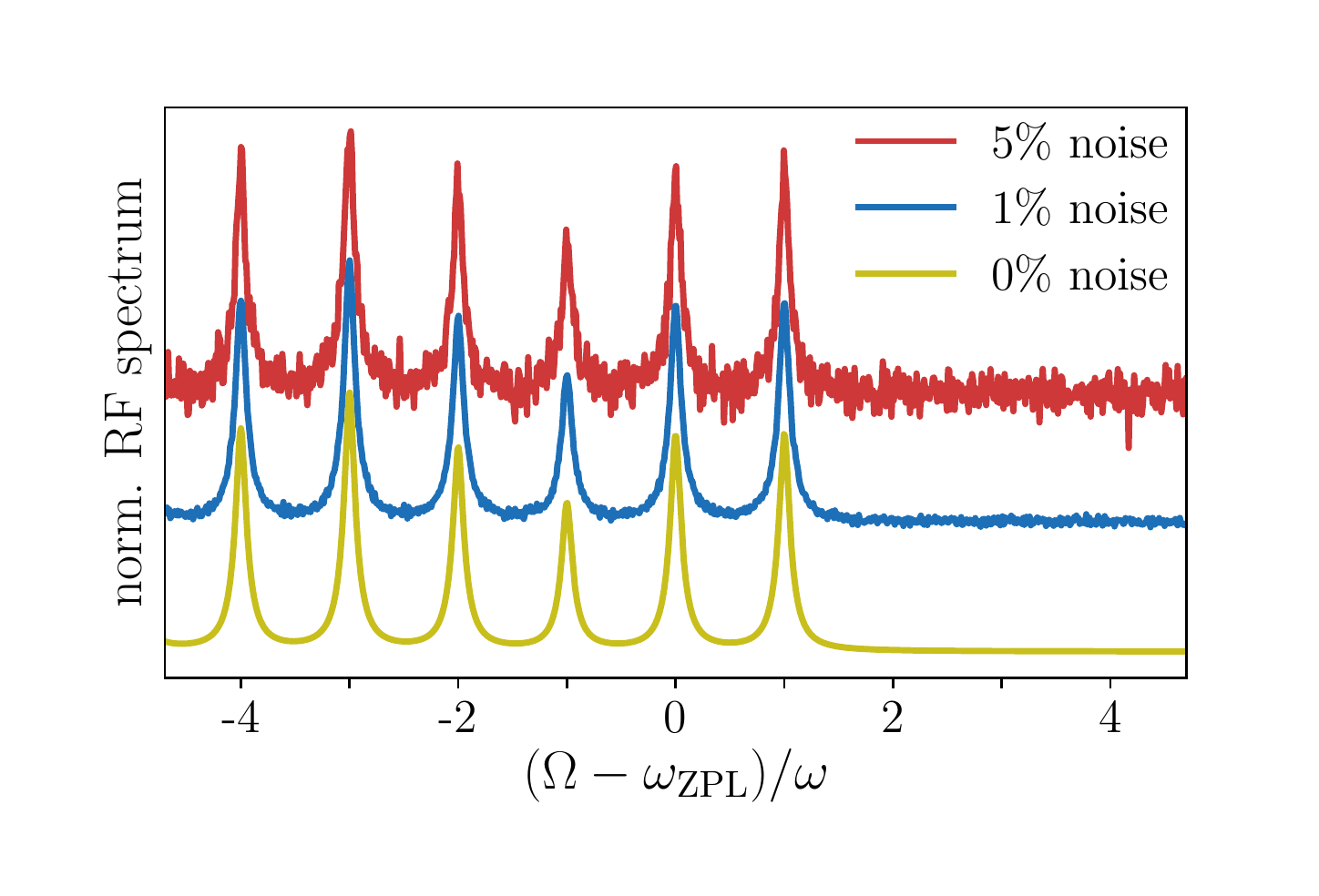}
		\caption{Full RF spectra from Eq.~\eqref{eq:S_total_complicated} for a symmetric superposition of initial phonon states $\ket{\text{0}}$ and $\ket{\text{1}}$. Additionally, Gaussian noise is added as a background with the standard deviation being a certain percentage of the largest peak in the spectrum, as indicated in the legend. The parameters are $\kappa=0$ (ZPL-excitation), $\gamma=\text{1.5}$ and $\gamma_\text{pd}=\gamma_\text{xd}=\Gamma=\text{0.05}\omega$.}
		\label{fig:noisyspectra}
	\end{figure}
\begin{figure}[h]
		\centering
		\includegraphics[width=0.5\linewidth]{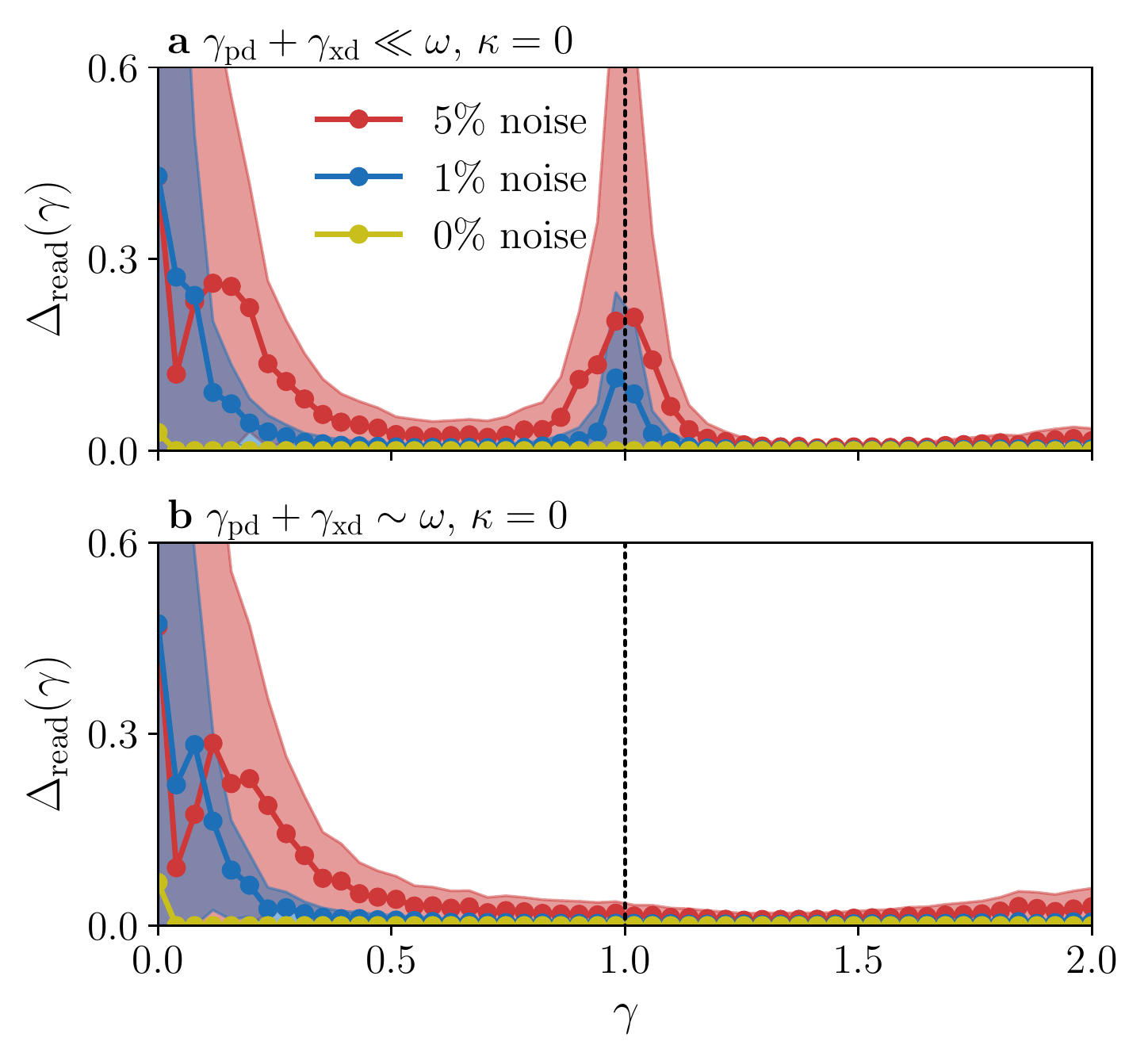}
		\caption{Readout error $\Delta_\text{read}$ vs. phonon coupling $\gamma$ obtained by fitting noisy spectra for ZPL-excitation with $\kappa=0$ and an initial symmetric superposition of the phonon states $\ket{\text{0}}$ and $\ket{\text{1}}$. The spectra are simulated using Eq.~\eqref{eq:S_total_complicated}, adding an additional background of Gaussian noise with the standard deviation being a certain percentage of the largest peak in the spectrum, ranging from 0\% (no noise) to 5\%, as indicated in the plot. As the fit function, the full simulation from Eq.~\eqref{eq:S_total_complicated} is used. The fits are performed for 50 different backgrounds, to make the results more independent of any specific realization of noise.
			The displayed values of $\Delta_\text{read}$ (colored dots with lines) and their respective standard deviations obtained from the fit (shaded colored areas) are the average values from these 50 realizations. (a) weak dissipation $\gamma_\text{pd}=\gamma_\text{xd}=\text{0.05}\omega$, (b) stronger dissipation $\gamma_\text{pd}=\gamma_\text{xd}=\text{0.5}\omega$. The spectrometer resolution is again $\Gamma=\text{0.05}\omega$.}
		\label{fig:noise_ZPL_10}
	\end{figure}

	So far we investigated how well we can obtain information on the initial phonon occupations using the approximation in Eq.~\eqref{eq:S_narrow}. This approximation relies on disregarding off-resonant photon-scattering processes. Thus, when for special values of the phonon coupling $\gamma$, certain transitions $\ket{i}_G\rightarrow\ket{i+\kappa}_X$ are strictly forbidden, no information on the initial occupation of $\ket{i}_G$ is contained in the approximative spectrum from Eq.~\eqref{eq:S_narrow}. This of course raises the question, why we do not simply use the full formula for the RF spectrum from Eq.~\eqref{eq:S_total_complicated}, which also contains off-resonant processes like $\ket{i}_G\rightarrow\ket{j}_X$ with $j\neq i+\kappa$. As we will see, using the full formula indeed permits to gain information on all phonon occupations, even at the special {\it blind} $\gamma$-values, where certain FC factors vanish. However, the previous discussion is still very useful to understand the underlying physics when going to a more realistic situation, where noise is present in the spectrum, which shall be fitted.\\
	Therefore, we consider the same superposition of $\ket{0}$ and $\ket{1}$ as before, but add noise to the spectrum in the form of uncorrelated white noise at discretized frequency points. Examples of such spectra are shown in Fig.~\ref{fig:noisyspectra} for different levels of noise. The standard deviation of this noise is chosen as a certain percentage of the largest peak intensity ranging from $0\%$ (no noise, yellow) to $5\%$ (red). The case of $1\%$ noise (blue) is comparable to typical experimental measurements \cite{weiss2021optomechanical}. A common origin for noise are for example dark counts in the single-photon detector~\cite{kang2003dark,shibata2015ultimate}.\\
	Now we directly use the full formula in Eq.~\eqref{eq:S_total_complicated} to extract the initial phonon occupations at different noise levels, as well as at different dephasing and decay rates. The resulting readout errors $\Delta_{\rm read}$ for the case of ZPL-excitation with $\kappa=0$ are plotted in Fig.~\ref{fig:noise_ZPL_10} as functions of the phonon coupling $\gamma$. The shaded areas show the $1\sigma$-intervals of the standard deviation for $\Delta_{\rm read}$ from the fitting procedure. Figure~\ref{fig:noise_ZPL_10}(a) shows the results for weak dissipation with $\gamma_{\rm pd}=\gamma_{\rm xd}=0.05\omega$, whereas in (b) we consider larger values with $\gamma_{\rm pd}=\gamma_{\rm xd}=0.5\omega$. The vertical dashed lines indicate the value of the phonon coupling $\gamma=1$, where the relevant FC factor $M_1^1$ for the transition $\ket{1}_G\rightarrow\ket{1}_X$ vanishes. In both cases, for vanishing noise (yellow dots with lines), the error $\Delta_{\rm read}$ rises towards $\gamma=0$ since in that case the phonons are decoupled from the TLS, such that no information on the phonon occupation can be obtained. In contrast to Fig.~\ref{fig:tracedistance10} however, there is no additional peak at the {\it blind} value of $\gamma=1$. The reason for this is of course, that the full spectrum in Eq.~\eqref{eq:S_total_complicated} contains the information on all initial phonon occupations due to off-resonant photon scattering, even if certain resonant transitions are forbidden.\\
	\noindent
	The situation changes drastically when adding noise to the studied spectrum (blue, red). To make the results more independent of any special realization of the added noise, we performed the fit for 50 different realizations and present here the average values. In both cases the readout error is largest around $\gamma=0$ up to roughly $\gamma\approx 0.2-0.5$. The range of large $\Delta_{\rm read}$ increases with increasing noise. The interpretation is of course the same as before. In the limit of $\gamma\rightarrow 0$, the phonons and the emitter are decoupled. For small $\gamma$, the PSBs are small compared to the ZPL and thus difficult to distinguish from the noise level, making the identification of the initial phonon occupations more challenging.\\
	In Fig.~\ref{fig:noise_ZPL_10}(a), which uses the same parameters as Fig.~\ref{fig:tracedistance10}, an additional peak is present at $\gamma=1$ in the cases of noisy RF spectra. For this value, the resonant transition $\ket{1}_G\rightarrow\ket{1}_X$ is forbidden. For $(\gamma_{\rm pd}+\gamma_{\rm xd})\ll\omega$, which is the case here, off-resonant scattering processes are suppressed, as it is exactly the parameter range, where the approximation from Eq.~\eqref{eq:S_amplitude_narrow} would be applicable. Therefore, the information on the initial phonon occupations, which is contained solely in these off-resonant scattering processes, is very vulnerable to noise. This leads to the same peak that we saw earlier in Fig.~\ref{fig:tracedistance10}(a). Also, the larger the noise, the larger the errors $\Delta_{\rm read}$ become.\\
	When we increase the dephasing and decay rates in (b), the peak at $\gamma=1$ disappears again for all noise values. This behavior can be explained as follows: The larger the dissipation in the system, the more likely off-resonant processes become, making the information contained in these processes more robust against noise. Thus, dissipation can actually help here to determine the phonon number statistics in the presence of noise.\\
	In total, we have seen that the same critical \textit{blind} values of the phonon coupling $\gamma$ emerge, when adding noise to the spectrum, even when using the full simulation from Eq.~\eqref{eq:S_total_complicated} for the fit. Therefore the approximation from Eq.~\eqref{eq:S_amplitude_narrow} can serve as a guidance to identify these critical values. The larger the contribution of noise to the spectra, the harder it is to identify the initial phonon number statistics in general. As discussed in Sec.~\ref{sec:detuning}, we can tune the critical \textit{blind} values for $\gamma$ by changing the detuning of the laser, which is a parameter that can be easily adjusted in an experiment. In addition, we have seen in this section that a faster dissipation of the emitter makes the identification of the phonon number statistics more robust against noise. 
	
	
	\section{Conclusion}
	In summary, we have  developed a method to read out the phonon number statistics from the sideband structure of resonance fluorescence spectra of a single-photon emitter. Starting from the well-established independent boson model for the coupling between a single phonon mode and an optically driven two-level system describing the emitter, we have developed analytical expressions for the detected light-scattering spectra in the limit of weak optical driving. We presented an in-depth discussion of the features of these spectra, including a discussion of the background due to inelastic scattering, present only in the case of non-vanishing pure dephasing of the emitter, and the semiclassical limit, which is in accordance with results known from quantum dots driven by surface acoustic waves. By using the common phonon quantum superposition state $(\ket{0}+\ket{1})/\sqrt{2}$ we then exemplarily demonstrated how the results can be used to read the number statistics of the phonon mode from the resonance fluorescence spectrum of the emitter. We have found that the readout routine is faulty when certain phonon-assisted transitions are forbidden due to vanishing Franck-Condon factors, depending on the phonon coupling strength. To resolve this issue, we have discussed how the dissipation of the emitter results in additional non-resonant scattering channels that can mitigate the missing Franck-Condon factors. Another strategy that we presented is to change the detuning between optical driving and emitter transition, which also results in variations of the Franck-Condon factors.
	
	In this work we restricted ourselves to a single phonon mode. The extension to multiple modes is straightforward, as long as the phonon-assisted excitation is unambiguous in the relevant range of laser detunings in the sense that for each line in the RF spectrum it is clear how many phonons from each mode are emitted or absorbed. This implies that there are no multi-phonon states with the same energy in the relevant spectral range. Otherwise, phonon wave-mixing becomes important, such that certain phonon-coherences do not average out in the time-integrated spectrum~\cite{weiss2021optomechanical}. This also prohibits a direct extension to systems with a continuous phonon spectrum.
		
	The described method used to read out the phonon number statistics from the optical scattering spectrum renders an important step in the development of a nanoscopic transducer between phonon and photon quantum states. The next step will be to consider more involved, e.g., time-dependent, spectroscopy signals, that also carry information on the initial phonon coherences, such that the entire phonon density matrix can be read out. This development would allow for a full hybrid integration between superconducting-qubit--based quantum acoustics and solid state quantum optics in the visible spectral regime. Such an integration will open a plethora of new possibilities to process, transmit, and store quantum information in solid state infrastructures.

	\medskip
	\textbf{Supporting Information} \par 
	Supporting Information is available from the Wiley Online Library or from the author.

	\medskip
	\textbf{Acknowledgements} \par 
	O.H. and D.W. acknowledge financial support by the Science Foundation Ireland (SFI) under\\Grant 18/RP/6236. D.G., P.M. and T.K. acknowledge financial support by the Alexander von Humobldt foundation (AvH) within the research group linkage programme funded by the German Federal Ministry of Education and Research (BMBF). We thank Matthias Wei\ss{} and Hubert Krenner for helpful discussions.
	
	\medskip
	\textbf{Conflict of Interest} \par 
	The authors declare no conflict of interest.
	
	\medskip
	
	%

\begin{thebibliography}{10}
	\providecommand{\url}[1]{\texttt{#1}}
	\providecommand{\urlprefix}{URL }
	
	\bibitem{uppu2021quantum}
	R.~Uppu, L.~Midolo, X.~Zhou, J.~Carolan, P.~Lodahl,
	\newblock \emph{Nat. Nanotechnol.} \textbf{2021}, \emph{16}, 12 1308.
	
	\bibitem{pezzagna2021quantum}
	S.~Pezzagna, J.~Meijer,
	\newblock \emph{Appl. Phys. Rev.} \textbf{2021}, \emph{8}, 1 011308.
	
	\bibitem{castelletto2020hexagonal}
	S.~Castelletto, F.~A. Inam, S.-I. Sato, A.~Boretti,
	\newblock \emph{Beilstein J. Nanotechnol.} \textbf{2020}, \emph{11}, 1 740.
	
	\bibitem{lodahl2015interfacing}
	P.~Lodahl, S.~Mahmoodian, S.~Stobbe,
	\newblock \emph{Rev. Mod. Phys.} \textbf{2015}, \emph{87}, 2 347.
	
	\bibitem{tan2015nonlinear}
	D.~T.~H. Tan, A.~M. Agarwal, L.~C. Kimerling,
	\newblock \emph{Laser Photonics Rev.} \textbf{2015}, \emph{9}, 3 294.
	
	\bibitem{xiong2021room}
	X.~Xiong, N.~Kongsuwan, Y.~Lai, C.~E. Png, L.~Wu, O.~Hess,
	\newblock \emph{Appl. Phys. Lett.} \textbf{2021}, \emph{118}, 13 130501.
	
	\bibitem{houck2008controlling}
	A.~A. Houck, J.~A. Schreier, B.~R. Johnson, J.~M. Chow, J.~Koch, J.~M.
	Gambetta, D.~I. Schuster, L.~Frunzio, M.~H. Devoret, S.~M. Girvin, R.~J.
	Schoelkopf,
	\newblock \emph{Phys. Rev. Lett.} \textbf{2008}, \emph{101}, 8 080502.
	
	\bibitem{xiang2013hybrid}
	Z.-L. Xiang, S.~Ashhab, J.~Q. You, F.~Nori,
	\newblock \emph{Rev. Mod. Phys.} \textbf{2013}, \emph{85}, 2 623.
	
	\bibitem{kurizki2015quantum}
	G.~Kurizki, P.~Bertet, Y.~Kubo, K.~M{\o}lmer, D.~Petrosyan, P.~Rabl,
	J.~Schmiedmayer,
	\newblock \emph{Proc. Natl. Acad. Sci. U.S.A.} \textbf{2015}, \emph{112}, 13
	3866.
	
	\bibitem{schuetz2015universal}
	M.~J.~A. Schuetz, E.~M. Kessler, G.~Giedke, L.~M.~K. Vandersypen, M.~D. Lukin,
	J.~I. Cirac,
	\newblock \emph{Phys. Rev. X} \textbf{2015}, \emph{5} 031031.
	
	\bibitem{lachance2019hybrid}
	D.~Lachance-Quirion, Y.~Tabuchi, A.~Gloppe, K.~Usami, Y.~Nakamura,
	\newblock \emph{Appl. Phys. Express} \textbf{2019}, \emph{12}, 7 070101.
	
	\bibitem{benito2020hybrid}
	M.~Benito, G.~Burkard,
	\newblock \emph{Appl. Phys. Lett.} \textbf{2020}, \emph{116}, 19 190502.
	
	\bibitem{delsing2019the}
	P.~Delsing, A.~N. Cleland, M.~J.~A. Schuetz, J.~Kn{\"o}rzer, G.~Giedke, J.~I.
	Cirac, K.~Srinivasan, M.~Wu, K.~C. Balram, C.~B{\"a}uerle, T.~Meunier,
	C.~J.~B. Ford, P.~V. Santos, E.~Cerda-M{\'e}ndez, H.~Wang, H.~J. Krenner,
	E.~D.~S. Nysten, M.~Wei{\ss}, G.~R. Nash, L.~Thevenard, C.~Gourdon,
	P.~Rovillain, M.~Marangolo, J.~Y. Duquesne, G.~Fischerauer, W.~Ruile,
	A.~Reiner, B.~Paschke, D.~Denysenko, D.~Volkmer, A.~Wixforth, H.~Bruus,
	M.~Wiklund, J.~Reboud, J.~M. Cooper, Y.~Q. Fu, M.~S. Brugger, F.~Rehfeldt,
	C.~Westerhausen,
	\newblock \emph{J. Phys. D: Appl. Phys.} \textbf{2019}, \emph{52}, 35 353001.
	
	\bibitem{lahaye2009nanomechanical}
	M.~D. LaHaye, J.~Suh, P.~M. Echternach, K.~C. Schwab, M.~L. Roukes,
	\newblock \emph{Nature} \textbf{2009}, \emph{459}, 7249 960.
	
	\bibitem{o2010quantum}
	A.~D. {O'Connell}, M.~Hofheinz, M.~Ansmann, R.~C. Bialczak, M.~Lenander,
	E.~Lucero, M.~Neeley, D.~Sank, H.~Wang, M.~Weides, J.~Wenner, J.~M. Martinis,
	A.~N. Cleland,
	\newblock \emph{Nature} \textbf{2010}, \emph{464}, 7289 697.
	
	\bibitem{mirhosseini2020superconducting}
	M.~Mirhosseini, A.~Sipahigil, M.~Kalaee, O.~Painter,
	\newblock \emph{Nature} \textbf{2020}, \emph{588}, 7839 599.
	
	\bibitem{gustafsson2014propagating}
	M.~V. Gustafsson, T.~Aref, A.~F. Kockum, M.~K. Ekstr{\"o}m, G.~Johansson,
	P.~Delsing,
	\newblock \emph{Science} \textbf{2014}, \emph{346}, 6206 207.
	
	\bibitem{chu2017quantum}
	Y.~Chu, P.~Kharel, W.~H. Renninger, L.~D. Burkhart, L.~Frunzio, P.~T. Rakich,
	R.~J. Schoelkopf,
	\newblock \emph{Science} \textbf{2017}, \emph{358}, 6360 199.
	
	\bibitem{hofheinz2009synthesizing}
	M.~Hofheinz, H.~Wang, M.~Ansmann, R.~C. Bialczak, E.~Lucero, M.~Neeley, A.~D.
	O'connell, D.~Sank, J.~Wenner, J.~M. Martinis, A.~N. Cleland,
	\newblock \emph{Nature} \textbf{2009}, \emph{459}, 7246 546.
	
	\bibitem{satzinger2018quantum}
	K.~J. Satzinger, Y.~P. Zhong, H.-S. Chang, G.~A. Peairs, A.~Bienfait, M.-H.
	Chou, A.~Y. Cleland, C.~R. Conner, {\'E}.~Dumur, J.~Grebel, I.~Gutierrez,
	B.~H. November, R.~G. Povey, S.~J. Whiteley, D.~D. Awschalom, D.~I. Schuster,
	A.~N. Cleland,
	\newblock \emph{Nature} \textbf{2018}, \emph{563}, 7733 661.
	
	\bibitem{chu2018creation}
	Y.~Chu, P.~Kharel, T.~Yoon, L.~Frunzio, P.~T. Rakich, R.~J. Schoelkopf,
	\newblock \emph{Nature} \textbf{2018}, \emph{563}, 7733 666.
	
	\bibitem{bild2022schr}
	M.~Bild, M.~Fadel, Y.~Yang, U.~von L{\"u}pke, P.~Martin, A.~Bruno, Y.~Chu,
	\newblock \emph{Science} \textbf{2023}, \emph{380}, 6642 274.
	
	\bibitem{schrinski2023macroscopic}
	B.~Schrinski, Y.~Yang, U.~von L{\"u}pke, M.~Bild, Y.~Chu, K.~Hornberger,
	S.~Nimmrichter, M.~Fadel,
	\newblock \emph{Phys. Rev. Lett.} \textbf{2023}, \emph{130}, 13 133604.
	
	\bibitem{choquer2022quantum}
	M.~Choquer, M.~Wei{\ss}, E.~D.~S. Nysten, M.~Lienhart, P.~Machnikowski,
	D.~Wigger, H.~J. Krenner, G.~Moody,
	\newblock \emph{IEEE Trans. Quantum Eng.} \textbf{2022}, \emph{3} 5100217.
	
	\bibitem{nysten2020hybrid}
	E.~D.~S. Nysten, A.~Rastelli, H.~J. Krenner,
	\newblock \emph{Appl. Phys. Lett.} \textbf{2020}, \emph{117}, 12 121106.
	
	\bibitem{wigger2021remote}
	D.~Wigger, K.~Gawarecki, P.~Machnikowski,
	\newblock \emph{Adv. Quantum Technol.} \textbf{2021}, \emph{4}, 4 2000128.
	
	\bibitem{imany2022quantum}
	P.~Imany, Z.~Wang, R.~A. DeCrescent, R.~C. Boutelle, C.~A. McDonald, T.~Autry,
	S.~Berweger, P.~Kabos, S.~W. Nam, R.~P. Mirin, K.~L. Silverman,
	\newblock \emph{Optica} \textbf{2022}, \emph{9}, 5 501.
	
	\bibitem{decrescent2022large}
	R.~A. DeCrescent, Z.~Wang, P.~Imany, R.~C. Boutelle, C.~A. McDonald, T.~Autry,
	J.~D. Teufel, S.~W. Nam, R.~P. Mirin, K.~L. Silverman,
	\newblock \emph{Phys. Rev. Appl.} \textbf{2022}, \emph{18}, 3 034067.
	
	\bibitem{bruggemann2012laser}
	C.~Br{\"u}ggemann, A.~V. Akimov, A.~V. Scherbakov, M.~Bombeck, C.~Schneider,
	S.~H{\"o}fling, A.~Forchel, D.~R. Yakovlev, M.~Bayer,
	\newblock \emph{Nat. Photonics} \textbf{2012}, \emph{6}, 1 30.
	
	\bibitem{czerniuk2017picosecond}
	T.~Czerniuk, D.~Wigger, A.~V. Akimov, C.~Schneider, M.~Kamp, S.~H{\"o}fling,
	D.~R. Yakovlev, T.~Kuhn, D.~E. Reiter, M.~Bayer,
	\newblock \emph{Phys. Rev. Lett.} \textbf{2017}, \emph{118}, 13 133901.
	
	\bibitem{krummheuer2002the}
	B.~Krummheuer, V.~M. Axt, T.~Kuhn,
	\newblock \emph{Phys. Rev. B} \textbf{2002}, \emph{65}, 19 195313.
	
	\bibitem{duke1965phonon}
	C.~B. Duke, G.~D. Mahan,
	\newblock \emph{Phys. Rev.} \textbf{1965}, \emph{139}, 6A A1965.
	
	\bibitem{weiss2021optomechanical}
	M.~Wei{\ss}, D.~Wigger, M.~N{\"a}gele, K.~M{\"u}ller, J.~J. Finley, T.~Kuhn,
	P.~Machnikowski, H.~J. Krenner,
	\newblock \emph{Optica} \textbf{2021}, \emph{8}, 3 291.
	
	\bibitem{wilson2004laser}
	I.~Wilson-Rae, P.~Zoller, A.~Imamo{\u{g}}lu,
	\newblock \emph{Phys. Rev. Lett.} \textbf{2004}, \emph{92}, 7 075507.
	
	\bibitem{groll2021controlling}
	D.~Groll, T.~Hahn, P.~Machnikowski, D.~Wigger, T.~Kuhn,
	\newblock \emph{Mater. Quantum Technol.} \textbf{2021}, \emph{1}, 1 015004.
	
	\bibitem{wigger2019phonon}
	D.~Wigger, R.~Schmidt, O.~Del Pozo-Zamudio, J.~A. Preu{\ss}, P.~Tonndorf,
	R.~Schneider, P.~Steeger, J.~Kern, Y.~Khodaei, J.~Sperling, S.~Michaelis~de
	Vasconcellos, R.~Bratschitsch, T.~Kuhn,
	\newblock \emph{2D Mater.} \textbf{2019}, \emph{6}, 3 035006.
	
	\bibitem{cui2016molecular}
	Y.~Cui, A.~Lauchner, A.~Manjavacas, F.~J. Garc{\'i}a~de Abajo, N.~J. Halas,
	P.~Nordlander,
	\newblock \emph{Nano Lett.} \textbf{2016}, \emph{16}, 10 6390.
	
	\bibitem{gilmore2005spin}
	J.~Gilmore, R.~H. McKenzie,
	\newblock \emph{J. Phys. Condens. Matter} \textbf{2005}, \emph{17}, 10 1735.
	
	\bibitem{braig2003vibrational}
	S.~Braig, K.~Flensberg,
	\newblock \emph{Phys. Rev. B} \textbf{2003}, \emph{68}, 20 205324.
	
	\bibitem{chen2005effects}
	Z.-Z. Chen, R.~L{\"u}, B.-F. Zhu,
	\newblock \emph{Phys. Rev. B} \textbf{2005}, \emph{71}, 16 165324.
	
	\bibitem{hahn2022photon}
	T.~Hahn, D.~Groll, H.~J. Krenner, T.~Kuhn, P.~Machnikowski, D.~Wigger,
	\newblock \emph{AVS Quantum Sci.} \textbf{2022}, \emph{4}, 1 011403.
	
	\bibitem{stauber2000electron}
	T.~Stauber, R.~Zimmermann, H.~Castella,
	\newblock \emph{Phys. Rev. B} \textbf{2000}, \emph{62}, 11 7336.
	
	\bibitem{forstner2003phonon}
	J.~F{\"o}rstner, C.~Weber, J.~Danckwerts, A.~Knorr,
	\newblock \emph{Phys. Rev. Lett.} \textbf{2003}, \emph{91}, 12 127401.
	
	\bibitem{hohenester2004quantum}
	U.~Hohenester, G.~Stadler,
	\newblock \emph{Phys. Rev. Lett.} \textbf{2004}, \emph{92}, 19 196801.
	
	\bibitem{nazir2016modelling}
	A.~Nazir, D.~P.~S. McCutcheon,
	\newblock \emph{J. Phys. Condens. Matter} \textbf{2016}, \emph{28}, 10 103002.
	
	\bibitem{carmele2019non}
	A.~Carmele, S.~Reitzenstein,
	\newblock \emph{Nanophotonics} \textbf{2019}, \emph{8}, 5 655.
	
	\bibitem{norambuena2016microscopic}
	A.~Norambuena, S.~A. Reyes, J.~Mej{\'\i}a-Lop{\'e}z, A.~Gali, J.~R. Maze,
	\newblock \emph{Phys. Rev. B} \textbf{2016}, \emph{94}, 13 134305.
	
	\bibitem{preuss2022resonant}
	J.~A. Preuss, D.~Groll, R.~Schmidt, T.~Hahn, P.~Machnikowski, R.~Bratschitsch,
	T.~Kuhn, S.~Michaelis De~Vasconcellos, D.~Wigger,
	\newblock \emph{Optica} \textbf{2022}, \emph{9}, 5 522.
	
	\bibitem{axt1999coherent}
	V.~M. Axt, M.~Herbst, T.~Kuhn,
	\newblock \emph{Superlattices Microstruct.} \textbf{1999}, \emph{26}, 2 117.
	
	\bibitem{reiter2011generation}
	D.~E. Reiter, D.~Wigger, V.~M. Axt, T.~Kuhn,
	\newblock \emph{Phys. Rev. B} \textbf{2011}, \emph{84}, 19 195327.
	
	\bibitem{alicki2004pure}
	R.~Alicki,
	\newblock \emph{Open Syst. Inf. Dyn.} \textbf{2004}, \emph{11}, 01 53.
	
	\bibitem{roszak2006path}
	K.~Roszak, P.~Machnikowski,
	\newblock \emph{Phys. Lett. A} \textbf{2006}, \emph{351}, 4-5 251.
	
	\bibitem{mahan2013many}
	G.~D. Mahan,
	\newblock \emph{Many-particle physics},
	\newblock Springer Science \& Business Media, New York, \textbf{2013}.
	
	\bibitem{glauber1963coherent}
	R.~J. Glauber,
	\newblock \emph{Phys. Rev.} \textbf{1963}, \emph{131} 2766.
	
	\bibitem{hohenester2007quantum}
	U.~Hohenester,
	\newblock \emph{J. Phys. B-At. Mol. Opt.} \textbf{2007}, \emph{40}, 11 S315.
	
	\bibitem{wigger2014energy}
	D.~Wigger, S.~L{\"u}ker, D.~E. Reiter, V.~M. Axt, P.~Machnikowski, T.~Kuhn,
	\newblock \emph{J. Phys. Condens. Matter} \textbf{2014}, \emph{26}, 35 355802.
	
	\bibitem{franck1926elementary}
	J.~Franck, E.~G. Dymond,
	\newblock \emph{Trans. Faraday Soc.} \textbf{1926}, \emph{21} 536.
	
	\bibitem{condon1926theory}
	E.~Condon,
	\newblock \emph{Phys. Rev.} \textbf{1926}, \emph{28}, 6 1182.
	
	\bibitem{maguire2019environmental}
	H.~Maguire, J.~Iles-Smith, A.~Nazir,
	\newblock \emph{Phys. Rev. Lett.} \textbf{2019}, \emph{123}, 9 093601.
	
	\bibitem{breuer2002theory}
	H.~P. Breuer, F.~Petruccione,
	\newblock \emph{{The Theory of Open Quantum Systems}},
	\newblock Oxford University Press, Oxford, \textbf{2002}.
	
	\bibitem{kohler1997floquet}
	S.~Kohler, T.~Dittrich, P.~H{\"a}nggi,
	\newblock \emph{Phys. Rev. E} \textbf{1997}, \emph{55}, 1 300.
	
	\bibitem{lazarides2014equilibrium}
	A.~Lazarides, A.~Das, R.~Moessner,
	\newblock \emph{Phys. Rev. E} \textbf{2014}, \emph{90}, 1 012110.
	
	\bibitem{morell2016high}
	N.~Morell, A.~Reserbat-Plantey, I.~Tsioutsios, K.~G. Sch{\"a}dler, F.~Dubin,
	F.~H.~L. Koppens, A.~Bachtold,
	\newblock \emph{Nano Lett.} \textbf{2016}, \emph{16}, 8 5102.
	
	\bibitem{tonndorf2015single}
	P.~Tonndorf, R.~Schmidt, R.~Schneider, J.~Kern, M.~Buscema, G.~A. Steele,
	A.~Castellanos-Gomez, H.~S.~J. van~der Zant, S.~Michaelis~de Vasconcellos,
	R.~Bratschitsch,
	\newblock \emph{Optica} \textbf{2015}, \emph{2}, 4 347.
	
	\bibitem{huttel2009carbon}
	A.~K. Huttel, G.~A. Steele, B.~Witkamp, M.~Poot, L.~P. Kouwenhoven, H.~S.~J.
	van~der Zant,
	\newblock \emph{Nano Lett.} \textbf{2009}, \emph{9}, 7 2547.
	
	\bibitem{laird2012high}
	F.~Laird, E. A .and~Pei, W.~Tang, G.~A. Steele, L.~P. Kouwenhoven,
	\newblock \emph{Nano Lett.} \textbf{2012}, \emph{12}, 1 193.
	
	\bibitem{he2017tunable}
	X.~He, N.~F. Hartmann, X.~Ma, Y.~Kim, R.~Ihly, J.~L. Blackburn, W.~Gao,
	J.~Kono, Y.~Yomogida, A.~Hirano, T.~Tanaka, H.~Kataura, H.~Htoon, S.~K.
	Doorn,
	\newblock \emph{Nat. Photonics} \textbf{2017}, \emph{11}, 9 577.
	
	\bibitem{eberly1977time}
	J.~H. Eberly, K.~Wodkiewicz,
	\newblock \emph{J. Opt. Soc. Am.} \textbf{1977}, \emph{67}, 9 1252.
	
	\bibitem{mollow1969power}
	B.~R. Mollow,
	\newblock \emph{Phys. Rev.} \textbf{1969}, \emph{188}, 5.
	
	\bibitem{wigger2021resonance}
	D.~Wigger, M.~Wei{\ss}, M.~Lienhart, K.~M{\"u}ller, J.~J. Finley, T.~Kuhn,
	H.~J. Krenner, P.~Machnikowski,
	\newblock \emph{Phys. Rev. Research} \textbf{2021}, \emph{3}, 3 033197.
	
	\bibitem{wu1975investigation}
	F.~Y. Wu, R.~E. Grove, S.~Ezekiel,
	\newblock \emph{Phys. Rev. Lett.} \textbf{1975}, \emph{35}, 21 1426.
	
	\bibitem{feynman2010quantum}
	R.~P. Feynman, A.~R. Hibbs, D.~F. Styer,
	\newblock Quantum mechanics and path integrals, \textbf{2010}.
	
	\bibitem{weiss2012quantum}
	U.~Weiss,
	\newblock \emph{Quantum dissipative systems},
	\newblock World Scientific, Singapore, \textbf{2012}.
	
	\bibitem{bogaczewicz2023resonance}
	R.~A. Bogaczewicz, P.~Machnikowski,
	\newblock \emph{{arXiv:2303.01531}} \textbf{2023}.
	
	\bibitem{dinu2020quantum}
	I.~V. Dinu, M.~{\c{T}}olea, P.~Gartner,
	\newblock \emph{Phys. Rev. B} \textbf{2020}, \emph{101}, 8 085304.
	
	\bibitem{NIST:DLMF}
	{\it NIST Digital Library of Mathematical Functions},
	\newblock \url{http://dlmf.nist.gov/}, Release 1.1.8 of 2022-12-15,
	\newblock \urlprefix\url{http://dlmf.nist.gov/},
	\newblock F.~W.~J. Olver, A.~B. {Olde Daalhuis}, D.~W. Lozier, B.~I. Schneider,
	R.~F. Boisvert, C.~W. Clark, B.~R. Miller, B.~V. Saunders, H.~S. Cohl, and
	M.~A. McClain, eds.
	
	\bibitem{weiss2018interfacing}
	M.~Wei{\ss}, H.~J. Krenner,
	\newblock \emph{J. Phys. D} \textbf{2018}, \emph{51}, 37 373001.
	
	\bibitem{vogele2020quantum}
	A.~Vogele, M.~M. Sonner, B.~Mayer, X.~Yuan, M.~Wei{\ss}, E.~D.~S. Nysten, S.~F.
	Covre~da Silva, A.~Rastelli, H.~J. Krenner,
	\newblock \emph{Adv. Quantum Technol.} \textbf{2020}, \emph{3}, 2 1900102.
	
	\bibitem{yeo2014strain}
	I.~Yeo, P.-L. de~Assis, A.~Gloppe, E.~Dupont-Ferrier, P.~Verlot, N.~S. Malik,
	E.~Dupuy, J.~Claudon, J.-M. G{\'e}rard, A.~Auff{\`e}ves, G.~Nogues,
	S.~Seidelin, J.-P. Poizat, O.~Arcizet, M.~Richard,
	\newblock \emph{Nat. Nanotechnol.} \textbf{2014}, \emph{9}, 2 106.
	
	\bibitem{more2006levenberg}
	J.~J. Mor{\'e},
	\newblock In \emph{Numerical Analysis: Proceedings of the Biennial Conference
		Held at Dundee, June 28--July 1, 1977}. Springer, \textbf{2006} 105--116.
	
	\bibitem{nielsen2002quantum}
	M.~A. Nielsen, I.~L. Chuang,
	\newblock \emph{Quantum Computation and Quantum Information: 10th Anniversary
		Edition},
	\newblock Cambridge University Press, Cambridge, \textbf{2010}.
	
	\bibitem{raimond2006exploring}
	J.-M. Raimond, S.~Haroche,
	\newblock \emph{Exploring the quantum},
	\newblock Oxford University Press, Oxford, \textbf{2006}.
	
	\bibitem{kang2003dark}
	Y.~Kang, H.~X. Lu, Y.-H. Lo, D.~S. Bethune, W.~P. Risk,
	\newblock \emph{Appl. Phys. Lett.} \textbf{2003}, \emph{83}, 14 2955.
	
	\bibitem{shibata2015ultimate}
	H.~Shibata, K.~Shimizu, H.~Takesue, Y.~Tokura,
	\newblock \emph{Opt. Lett.} \textbf{2015}, \emph{40}, 14 3428.
	
\end{thebibliography}

\end{document}


\pagestyle{fancy}
	\rhead{\includegraphics[width=2.5cm]{VCH-logo.png}}

	\title{Supporting Information: How to read out the phonon number statistics via resonance fluorescence spectroscopy of a single-photon emitter}
	
	\maketitle

	
	\author{Daniel Groll*}
	\author{Fabian Paschen}
	\author{Pawe\l{} Machnikowski}
	\author{Ortwin Hess}
	\author{Daniel Wigger}
	\author{Tilmann Kuhn*}

	\begin{affiliations}
		Daniel Groll, Fabian Paschen, Prof. Tilmann Kuhn\\
		Institute of Solid State Theory, University of M\"unster, 48149 M\"unster, Germany\\
		
		Prof. Pawe\l{} Machnikowski\\
		Institute of Theoretical Physics, Wroc\l{}aw University of Science and Technology, 50-370 Wroc\l{}aw, Poland\\
		
		Prof. Ortwin Hess, Dr. Daniel Wigger\\
		School of Physics, Trinity College Dublin, Dublin 2, Ireland\\
		
		Prof. Ortwin Hess\\
		CRANN Institute and Advanced Materials and Bioengineering Research (AMBER), Trinity College Dublin, Dublin 2, Ireland\\
		
		Email Addresses: daniel.groll@uni-muenster.de, tilmann.kuhn@uni-muenster.de	
	\end{affiliations}
	
		\section{Time-integrated spectrum from time-dependent spectrum}\label{app:stationary_spectrum}
	In the calculation of resonance fluorescence (RF) signals we are interested in the time-integrated spectrum, detected by a spectrometer. This can be modeled using the time-dependent spectrum~\cite{eberly1977time, groll2021controlling}
	\begin{align}
	&S(t,\Omega;\Gamma)=2\Gamma^2\text{Re}\left[\int\limits_{0}^{\infty}\text{d}\tau\,\int\limits_{-\infty}^t\text{d}t'\,e^{-\Gamma\tau}e^{-2\Gamma(t-t')}e^{-i\Omega\tau}G(t',t'-\tau)\right]\,,
	\end{align}
	where $\Gamma$ is the resolution of the spectrometer, prototypically modeled as a Fabry-Perot interferometer with an adjustable resonance frequency $\Omega$ in front of a photon counter (see Fig.~2 in the main text). The time-dependent spectrum is determined by the correlation function~\cite{mollow1969power}
	\begin{equation}\label{eq:G_t_tau}
	G(t',t'-\tau)=\expval{X^{\dagger}(t')X(t'-\tau)}\,.
	\end{equation}
	The time-integrated spectrum can be obtained from averaging the time-dependent spectrum over $t$, such that
	\begin{align}\label{eq:S_step}
	S(\Omega;\Gamma)&=\lim\limits_{T\rightarrow\infty}\frac{2\Gamma^2}{T}\text{Re}\left[\rule{0cm}{7mm}\right.\int\limits_{t_0}^{t_0+T}\text{d}t\,\int\limits_{0}^{\infty}\text{d}\tau\,\int\limits_{-\infty}^t\text{d}t'\,   e^{-\Gamma\tau}e^{-2\Gamma(t-t')}e^{-i\Omega\tau}G(t',t'-\tau)\left.\rule{0cm}{7mm}\right]\notag\\
	&=2\Gamma^2\text{Re}\left[ \int\limits_{0}^{\infty}\text{d}\tau\,\int\limits_{0}^{\infty}\text{d}t''\, e^{-\Gamma\tau}e^{-2\Gamma t''}e^{-i\Omega\tau}\overline{G}(\tau) \right]
	\end{align}
	with $t''=t-t'$ and
	\begin{align}
	\overline{G}(\tau)&=\lim\limits_{T\rightarrow\infty}\frac{1}{T}\int\limits_{t_0}^{t_0+T}\text{d}t\,G(t-t'',t-t''-\tau)=\lim\limits_{T\rightarrow\infty}\frac{1}{T}\int\limits_{t_0}^{t_0+T-t''}\text{d}t\,G(t,t-\tau)
	\end{align}
	being the average of the correlation function with respect to $t$. In the last step, we neglected any contributions to the correlation function in the time interval from $t_0-t''$ to $t_0$, assuming that the TLS is initially in the ground state and that the optical driving is turned on at time $t_0$. Note that this time-averaged correlation function is independent of $t''$ for any finite $t''$. In Eq.~\eqref{eq:S_step}, $t''$ is integrated up to $t''=\infty$ but since the contributions to $S(\Omega;\Gamma)$ are suppressed exponentially with increasing $t''$, it is valid to assume that $\overline{G}$ is independent of $t''$ under the integral in Eq.~\eqref{eq:S_step}. With the same argument (for $\tau$) we can express the time-averaged correlation function via
	\begin{equation}\label{eq:G_bar}
	\overline{G}(\tau)=\lim\limits_{T\rightarrow\infty}\frac{1}{T}\int\limits_{t_0}^{t_0+T}\text{d}t\,G(t+\tau,t)\,.
	\end{equation}
	Finally, the time-integrated spectrum reads
	\begin{equation}\label{eq:S_RF}
	S(\Omega;\Gamma)=\Gamma \text{Re}\left[\int\limits_0^{\infty}\text{d}\tau\,e^{-(\Gamma+i\Omega)\tau}\overline{G}(\tau)\right]\,.
	\end{equation}
	\section{Calculation of the time-averaged correlation function}\label{app:aver_corr_func}
	Our goal is to calculate the time-averaged correlation function from Eq.~\eqref{eq:G_bar}, where the TLS is initially in the ground state $\rho(t_0)=\ket{G}\bra{G}\rho^G$ and $\rho^G$ is the initial phonon state. As discussed in Sec.~3 of the main text, there are two contributions to the correlation function
	\begin{equation}
	G(t+\tau,t)=\Tr(X^{\dagger}\mathcal{V}(t+\tau,t)\left\lbrace X\mathcal{V}(t,t_0)\left[\rho(t_0)\right]\right\rbrace)\,,
	\end{equation}
	namely the quantities $\bra{X}\mathcal{V}(t,t_0)\left[\rho(t_0)\right]\ket{X}$ and $\bra{X}\mathcal{V}(t,t_0)\left[\rho(t_0)\right]\ket{G}$, i.e., the excited state and coherence contribution. These are determined in the following in second order with respect to the driving field, employing the perturbation expansion of the time evolution super-operator $\mathcal{V}(t,t_0)$ from Eq.~(18) in the main text.
	\subsection{Excited state}\label{sec:XX}
	First we focus on the subspace of the excited TLS, i.e., we want to determine $\bra{X}\rho(t)\ket{X}$ in second order with respect to the driving field. Since the free time evolution [Eq.~(17b), main text] does not generate occupation of $\ket{X}$, but only destroys it via $\mathcal{D}_{\mathrm{xd}}$, Eq.~(18) in the main text reduces to the last term, which reads
	\begin{equation}
	\bra{X}\rho(t)\ket{X}=\int\limits_{t_0}^t\text{d}\tau_1\,\int\limits_{t_0}^{\tau_1}\text{d}\tau_2\,\bra{X}\left[e^{\mathcal{L}_0(t-\tau_1)}\mathcal{L}_I(\tau_1)e^{\mathcal{L}_0(\tau_1-\tau_2)}\mathcal{L}_I(\tau_2)e^{\mathcal{L}_0(\tau_2-t_0)}\rho(t_0)\right]\ket{X}\,.
	\end{equation}
	With 
	\begin{subequations}\begin{align}
		\mathcal{D}_{\mathrm{pd}}(\ket{G}\bra{G})&= \mathcal{D}_{\mathrm{xd}}(\ket{G}\bra{G}) =\mathcal{D}_{\mathrm{pd}}(\ket{X}\bra{X})=0\,,\label{eq:trivial_D}\\
		\mathcal{D}_{\mathrm{xd}}(\ket{X}\bra{X})&=-\gamma_{\rm xd}\ket{X}\bra{X}+\gamma_{\rm xd}\ket{G}\bra{G}\,,\label{eq:D_x}\\
		\mathcal{D}_{\mathrm{j}}(\ket{N}\bra{M})&=-\frac{\gamma_{\rm j}}{2}\ket{N}\bra{M}\,,\quad \textrm{j}=\textrm{pd, xd},\quad N,M=X,G\ (N\neq M)\,,\label{eq:D_coherence}
		\end{align}\end{subequations}
	we can determine the action of the free time evolution due to $\mathcal{L}_0$ on the different time intervals. From $t_0$ to $\tau_2$ the TLS is in its ground state, rendering the action of $\mathcal{D}_{\mathrm{pd/xd}}$ in $\exp[\mathcal{L}_0(\tau_2-t_0)]$ trivial according to Eq.~\eqref{eq:trivial_D}, leading to
	\begin{equation}
	e^{\mathcal{L}_0(\tau_2-t_0)}\rho(t_0)=e^{-\frac{i}{\hbar}H_0(\tau_2-t_0)}\rho(t_0)e^{\frac{i}{\hbar}H_0(\tau_2-t_0)}\,.
	\end{equation}
	The action of $\mathcal{L}_I(\tau_2)$, i.e., the commutator with $H_I(\tau_2)$, yields coherence contributions of the form $\ket{G}\bra{X}$ and $\ket{X}\bra{G}$. In addition to the free time evolution with $H_0$ we thus get an exponential damping due to Eq.~\eqref{eq:D_coherence}
	\begin{equation}
	e^{\mathcal{L}_0(\tau_1-\tau_2)}\mathcal{L}_I(\tau_2)e^{\mathcal{L}_0(\tau_2-t_0)}\rho(t_0)=e^{-\frac{\gamma_{\rm pd}+\gamma_{\rm xd}}{2}(\tau_1-\tau_2)}e^{-\frac{i}{\hbar}H_0(\tau_1-\tau_2)}\left[\mathcal{L}_I(\tau_2)e^{\mathcal{L}_0(\tau_2-t_0)}\rho(t_0)\right]e^{\frac{i}{\hbar}H_0(\tau_1-\tau_2)}\,.
	\end{equation}
	The action of $\mathcal{L}_I(\tau_1)$ yields ground state contributions proportional to $\ket{G}\bra{G}$ and excited state contributions proportional to $\ket{X}\bra{X}$. Only the latter contribute here and they evolve in time with $H_0$ as well as decay exponentially with the rate $\gamma_{\rm xd}$ due to Eq.~\eqref{eq:D_x} between times $\tau_1$ and $t$. In total, after writing out the action of $\mathcal{L}_I$, this yields
	\begin{align}
	\bra{X}\rho(t)\ket{X}=\int\limits_{t_0}^t\text{d}\tau_1\,\int\limits_{t_0}^{\tau_1}\text{d}\tau_2&\, e^{-\gamma_{\rm xd}(t-\tau_1)-\frac{\gamma_{\rm pd}+\gamma_{\rm xd}}{2}(\tau_1-\tau_2)}\frac{1}{4\hbar^2}\mathcal{E}(\tau_1)\mathcal{E}^*(\tau_2)\notag\\
	&\times \bra{X}e^{-\frac{i}{\hbar}H_0(t-\tau_1)}X^{\dagger}e^{-\frac{i}{\hbar}H_0(\tau_1-t_0)}\rho(t_0)e^{\frac{i}{\hbar}H_0(\tau_2-t_0)}Xe^{\frac{i}{\hbar}H_0(t-\tau_2)}\ket{X}+h.c.
	\end{align}
	We can symmetrize this expression by removing the time-ordering. This is done by switching $\tau_2$ and $\tau_1$ in the hermitian conjugate ($h.c.$) part of the last expression and using
	\begin{equation}
	\int\limits_{t_0}^t\dd\tau_2\,\int\limits_{t_0}^{\tau_2}\dd\tau_1\,...=\int\limits_{t_0}^t\dd\tau_1\,\int\limits_{\tau_1}^t\dd\tau_2\,...
	\end{equation}
	to obtain
	\begin{align}
	\bra{X}\rho(t)\ket{X}=\int\limits_{t_0}^t\text{d}\tau_1\,\int\limits_{t_0}^t\text{d}\tau_2&\,e^{-\frac{\gamma_{\rm xd}}{2}(2t-\tau_2-\tau_1)}e^{-\frac{\gamma_{\rm pd}}{2}|\tau_1-\tau_2|}\frac{1}{4\hbar^2}\mathcal{E}(\tau_1)\mathcal{E}^*(\tau_2)\notag\\
	&\times \bra{X}e^{-\frac{i}{\hbar}H_0(t-\tau_1)}X^{\dagger}e^{-\frac{i}{\hbar}H_0(\tau_1-t_0)}\rho(t_0)e^{\frac{i}{\hbar}H_0(\tau_2-t_0)}Xe^{\frac{i}{\hbar}H_0(t-\tau_2)}\ket{X}\,.
	\end{align}
	This expression is still an operator on the phonon subspace. We now want to investigate its matrix elements in the appropriate energy eigenbasis using the properties of $H_0$, discussed in Sec.~2 in the main text. Projecting $\bra{m}B_+$ from the left and $B_-\ket{n}$ from the right, we can calculate
	\begin{align}
	\braind{X}{m}\rho(t)\ket{n}_X&=\int\limits_{t_0}^t\text{d}\tau_1\,\int\limits_{t_0}^t\text{d}\tau_2\,e^{-\frac{\gamma_{\rm xd}}{2}(2t-\tau_2-\tau_1)}e^{-\frac{\gamma_{\rm pd}}{2}|\tau_1-\tau_2|}\frac{1}{4\hbar^2}\mathcal{E}(\tau_1)\mathcal{E}^*(\tau_2)e^{-i\omega_{\rm ZPL}(\tau_2-\tau_1)}\notag\\
	&\qquad\times\sum_{p,q} e^{-im\omega(t-\tau_1)}\bra{m}B_+\ket{p}e^{-ip\omega(\tau_1-t_0)}\bra{p}\rho^G\ket{q}e^{iq\omega(\tau_2-t_0)}\bra{q}B_-\ket{n}e^{in\omega(t-\tau_2)}\notag\\
	&=\sum_{p,q} M_m^p M_n^{q*} \rho^G_{p,q}I_{m,p,q,n}(t,t_0)\,,\label{eq:rho_X_appendix}
	\end{align}
	where we introduced the Franck-Condon (FC) factors $\bra{m}B_+\ket{n}=M_{m}^n$~\cite{franck1926elementary, condon1926theory, maguire2019environmental}.  For the case of cw-driving, the double integral $I_{m,p,q,n}(t,t_0)$ can be solved by elementary integration. In RF experiments one is usually interested in the stationary signal, not the transient (of the TLS), such that we can consider $(t-t_0)\gg\gamma_{\rm xd}^{-1}$, leading to
	\begin{align}
	I_{m,p,q,n}(t,t_0)&=e^{-i\omega(p-q)(t-t_0)}\frac{|\mathcal{E}_0|^2}{4\hbar^2}\frac{\gamma_{\rm pd}+\gamma_{\rm xd}+i\omega(m-n-p+q)}{\gamma_{\rm xd}+i\omega(m-n-p+q)}\notag\\
	&\qquad\times\left[\frac{\gamma_{\rm pd}+\gamma_{\rm xd}}{2}+i\omega(\kappa+q-n)\right]^{-1}\left[\frac{\gamma_{\rm pd}+\gamma_{\rm xd}}{2}-i\omega(\kappa+p-m)\right]^{-1}\,.
	\end{align}
	
	\subsection{Coherence}\label{sec:XG}
	Next, we take a look at the coherence contribution $\bra{X}\rho(t)\ket{G}$ in an analog fashion to the case $\bra{X}\rho(t)\ket{X}$. The only non-vanishing contribution up to second order in $H_I$ follows from the first-order term in Eq.~(18) in the main text and is given by
	\begin{align}
	\bra{X}\rho(t)\ket{G}&=\bra{X}\left[\ \int\limits_{t_0}^t\text{d}\tau_1\,e^{\mathcal{L}_0(t-\tau_1)}\mathcal{L}_I(\tau_1)e^{\mathcal{L}_0(\tau_1-t_0)}\rho(t_0)\right]\ket{G}\notag\\
	&=-\frac{i}{2\hbar}\int\limits_{t_0}^t\text{d}\tau_1\,\mathcal{E}(\tau_1)e^{-\frac{\gamma_{\rm xd}+\gamma_{\rm pd}}{2}(t-\tau_1)}\bra{X}e^{-\frac{i}{\hbar}H_0(t-\tau_1)}X^{\dagger}e^{-\frac{i}{\hbar}H_0(\tau_1-t_0)}\rho(t_0)e^{\frac{i}{\hbar}H_0(t-t_0)}\ket{G}\,.
	\end{align}
	Using the energy eigenstates of $H_0$, we can write
	\begin{align}\label{eq:coherence_integral}
	\braind{X}{m}\rho(t)\ket{n}_G&=-\frac{i\mathcal{E}_0}{2\hbar}e^{-i\omega_{\rm ZPL}t}\int\limits_{t_0}^t\text{d}\tau_1\,e^{-\frac{\gamma_{\rm xd}+\gamma_{\rm pd}}{2}(t-\tau_1)}e^{-i\kappa\omega\tau_1}\notag\\
	&\qquad\times\sum_pe^{-im\omega(t-\tau_1)}M_m^pe^{-ip\omega(\tau_1-t_0)}\rho^G_{p,n}e^{in\omega(t-t_0)}\,.
	\end{align}
	Upon integration, this yields
	\begin{align}\label{eq:coherence_integral_result}
	\braind{X}{m}\rho(t)\ket{n}_G&=-\frac{i\mathcal{E}_0}{2\hbar}e^{-i\omega_{\rm ZPL}t}\sum_p\rho^G_{p,n}e^{in\omega(t-t_0)}M_m^p\left[\frac{\gamma_{\rm pd}+\gamma_{\rm xd}}{2}-i\omega(\kappa-m+p)\right]^{-1}\notag\\
	&\qquad\times \Big[e^{-i\kappa\omega t}e^{-ip\omega(t-t_0)}-e^{-\frac{\gamma_{\rm xd}+\gamma_{\rm pd}}{2}(t-t_0)}e^{-i\kappa\omega t_0}e^{-im\omega(t-t_0)}\Big]\,.
	\end{align}
	\subsection{Calculation of the two contributions to the RF spectrum}
	In this section we combine the calculations of the excited state and coherence part of the density matrix from the previous sections to evaluate the correlation function in Eq.~\eqref{eq:G_t_tau}
	to the lowest non-vanishing, i.e., quadratic, order in the driving field. We can re-write the correlation function as
	\begin{align}
	G(t+\tau,t)&=\text{Tr}\left(X^{\dagger}\mathcal{V}(t+\tau,t)\left\lbrace X\rho(t)\right\rbrace\right)\\
	&=\text{Tr}\left(\ket{X}\bra{G}\mathcal{V}(t+\tau,t)\left\lbrace \ket{G}\bra{X}\rho(t)\qty(\ket{G}\bra{G}+\ket{X}\bra{X})\right\rbrace\right)\notag\\
	&=\text{Tr}\left(\ket{X}\bra{G}\mathcal{V}(t+\tau,t)\left\lbrace\ket{G}\bra{X}\rho(t)\ket{G}\bra{G}\right\rbrace\right)+\text{Tr}\left(\ket{X}\bra{G}\mathcal{V}(t+\tau,t)\left\lbrace\ket{G}\bra{X}\rho(t)\ket{X}\bra{X}\right\rbrace\right)\notag\,.
	\end{align}
	The first contribution $\sim\bra{X}\rho(t)\ket{G}$ is linear in the driving field for the propagation from $t_0$ to $t$ and thus has to be linear for the propagation from $t$ to $t+\tau$, which yields a quadratic contribution in total. The second contribution is already quadratic in the driving field from $t_0$ to $t$, such that we can restrict ourselves to field-free propagation from $t$ to $t+\tau$~\cite{weiss2021optomechanical}.
	\subsubsection{Coherence contribution}
	The first contribution can be evaluated as (dropping the trace for the moment)
	\begin{align}
	&\bra{G}\left[\mathcal{V}(t+\tau,t)\left\lbrace\ket{G}\bra{X}\rho(t)\ket{G}\bra{G}\right\rbrace\right]\ket{X}\notag\\
	&=\bra{G}\left[\int\limits_{t}^{t+\tau}\text{d}\tau_1\,e^{\mathcal{L}_0(t+\tau-\tau_1)}\mathcal{L}_I(\tau_1) e^{\mathcal{L}_0(\tau_1-t)}\ket{G}\bra{X}\rho(t)\ket{G}\bra{G}\right.\biggr]\ket{X}\notag\\
	&=\frac{i}{2\hbar}\int\limits_{t}^{t+\tau}\text{d}\tau_1\,\mathcal{E}^*(\tau_1)e^{-\frac{\gamma_{\rm xd}+\gamma_{\rm pd}}{2}(t+\tau-\tau_1)}\bra{G}e^{-\frac{i}{\hbar}H_0\tau}\ket{G}\bra{X}\rho(t)\ket{G}\bra{G}e^{\frac{i}{\hbar}H_0(\tau_1-t)}\ket{G}\bra{X}e^{\frac{i}{\hbar}H_0(t+\tau-\tau_1)}\ket{X}\,.
	\end{align}
	The calculation is completely analog to that of the coherence in the previous section, leading to 
	\begin{align}
	\braind{G}{m}\left[\mathcal{V}(t+\tau,t)\left\lbrace\ket{G}\bra{X}\rho(t)\ket{G}\bra{G}\right\rbrace\right]\ket{n}_X&=\frac{i\mathcal{E}_0^*}{2\hbar}e^{i\omega_{\rm ZPL}(t+\tau)}\int\limits_{t}^{t+\tau}\text{d}\tau_1\,e^{i\kappa\omega\tau_1}e^{-\frac{\gamma_{\rm xd}+\gamma_{\rm pd}}{2}(t+\tau-\tau_1)}e^{-im\omega\tau} \notag\\
	&\qquad\times\sum_p\underbrace{\bra{m}\bra{X}\rho(t)\ket{G}\ket{p}}_{\bar{\rho}^{G*}_{p,m}}e^{ip\omega(\tau_1-t)}M_n^{p*}e^{in\omega(t+\tau-\tau_1)}\,.
	\end{align}
	This is the complex conjugate of Eq.~\eqref{eq:coherence_integral} with the replacements $t\rightarrow t+\tau$, $t_0\rightarrow t$, $\rho^G_{p,n}\rightarrow \bar{\rho}^G_{p,n}$, and $m\leftrightarrow n$. Thus, the result reads
	\begin{align}
	\braind{G}{m}\left[\mathcal{V}(t+\tau,t)\left\lbrace\ket{G}\bra{X}\rho(t)\ket{G}\bra{G}\right\rbrace\right]\ket{n}_X&=\frac{i\mathcal{E}^*_0}{2\hbar}e^{i\omega_{\rm ZPL}(t+\tau)}\sum_p\bar{\rho}^{G*}_{p,m}e^{-im\omega\tau}M_n^{p*}\notag\\
	&\qquad\times \left[\frac{\gamma_{\rm pd}+\gamma_{\rm xd}}{2}+i\omega(\kappa-n+p)\right]^{-1}\notag\\
	&\qquad\times \left[e^{i\kappa\omega (t+\tau)}e^{ip\omega\tau}-e^{-\frac{\gamma_{\rm xd}+\gamma_{\rm pd}}{2}\tau}e^{i\kappa\omega t}e^{in\omega\tau}\right]\,.
	\end{align}
	Using Eq.~\eqref{eq:coherence_integral_result} and neglecting the transient of the TLS, i.e., considering $(t-t_0)\gg \gamma_{\rm xd}^{-1}$, we obtain
	\begin{align}
	\braind{G}{m}\left[\mathcal{V}(t+\tau,t)\left\lbrace\ket{G}\bra{X}\rho(t)\ket{G}\bra{G}\right\rbrace\right]\ket{n}_X&=\frac{|\mathcal{E}_0|^2}{4\hbar^2}e^{i\omega_{\rm ZPL}\tau}\sum_{p,q,r}e^{-i(m-n)\omega\tau}M_q^{m*}M_n^{p*}M_q^r\rho_{r,p}^G\notag\\
	&\qquad\times\left[\frac{\gamma_{\rm pd}+\gamma_{\rm xd}}{2}+i\omega(\kappa-n+p)\right]^{-1}\left[e^{i(\kappa+p-n)\omega \tau}-e^{-\frac{\gamma_{\rm xd}+\gamma_{\rm pd}}{2}\tau}\right]\notag\\
	&\qquad\times \left[\frac{\gamma_{\rm pd}+\gamma_{\rm xd}}{2}-i\omega(\kappa-q+r)\right]^{-1}e^{i(p-r)\omega(t-t_0)}\,,\label{eq:coh_contr}
	\end{align}
	where we made use of
	\begin{align}
	\bar{\rho}^{G*}_{p,m}&=\bra{m}\bra{X}\rho(t)\ket{G}\ket{p}=\sum_q\bra{m}B_-\ket{q}\bra{q}B_+\bra{X}\rho(t)\ket{G}\ket{p}=\sum_qM_q^{m*}\braind{X}{q}\rho(t)\ket{p}_G\,.
	\end{align}
	This result can be inserted into
	\begin{align}
	\text{Tr}\left[\ket{X}\bra{G}\mathcal{V}(t+\tau,t)\left\lbrace\ket{G}\bra{X}\rho(t)\ket{G}\bra{G}\right\rbrace\right]=\sum_{m,n}M_n^m \braind{G}{m}\left[\mathcal{V}(t+\tau,t)\left\lbrace\ket{G}\bra{X}\rho(t)\ket{G}\bra{G}\right\rbrace\right]\ket{n}_X .
	\end{align}
	\subsubsection{Occupation contribution}
	The second contribution is easier to evaluate since the propagation from $t$ to $t+\tau$ is here of zeroth order in the driving field. Thus we get
	\begin{align}
	\bra{G}\left[\mathcal{V}(t+\tau,t)\left\lbrace\ket{G}\bra{X}\rho(t)\ket{X}\bra{X}\right\rbrace\right]\ket{X}&=\bra{G}\left[e^{\mathcal{L}_0\tau}\ket{G}\bra{X}\rho(t)\ket{X}\bra{X}\right]\ket{X}\notag\\
	&=e^{-\frac{\gamma_{\rm pd}+\gamma_{\rm xd}}{2}\tau}\bra{G}e^{-\frac{i}{\hbar}H_0\tau}\ket{G}\bra{X}\rho(t)\ket{X}\bra{X}e^{\frac{i}{\hbar}H_0\tau}\ket{X}\,,
	\end{align}
	which leads to
	\begin{align}
	\braind{G}{m}\left[\mathcal{V}(t+\tau,t)\left\lbrace\ket{G}\bra{X}\rho(t)\ket{X}\bra{X}\right\rbrace\right]\ket{n}_X&=e^{i\omega_{\rm ZPL}\tau}e^{-\frac{\gamma_{\rm pd}+\gamma_{\rm xd}}{2}\tau}e^{-i(m-n)\omega\tau}\bra{m}\bra{X}\rho(t)\ket{X}B_-\ket{n}\label{eq:occ_contr}\\
	&=\sum_q M_q^{m*}e^{i\omega_{\rm ZPL}\tau}e^{-\frac{\gamma_{\rm pd}+\gamma_{\rm xd}}{2}\tau}e^{-i(m-n)\omega\tau}\braind{X}{q}\rho(t)\ket{n}_X\notag\\
	&=\sum_{q,p,r} M_q^{m*}M_q^pM_n^{r*}\rho_{p,r}^Ge^{i\omega_{\rm ZPL}\tau}e^{-\frac{\gamma_{\rm pd}+\gamma_{\rm xd}}{2}\tau}e^{-i(m-n)\omega\tau}I_{q,p,r,n}(t,t_0)\notag\,.
	\end{align}
	This result can be inserted into
	\begin{align}
	\text{Tr}\left[\ket{X}\bra{G}\mathcal{V}(t+\tau,t)\left\lbrace\ket{G}\bra{X}\rho(t)\ket{X}\bra{X}\right\rbrace\right]=\sum_{m,n}M_n^m \braind{G}{m}\left[\mathcal{V}(t+\tau,t)\left\lbrace\ket{G}\bra{X}\rho(t)\ket{X}\bra{X}\right\rbrace\right]\ket{n}_X .
	\end{align}
	\subsection{Time-averaged correlation function and spectrum}
	For the RF signal in Eq.~\eqref{eq:S_RF}, we only need to calculate the time-averaged correlation function $\overline{G}(\tau)$ in Eq.~\eqref{eq:G_bar}, which enforces $p=r$ under the sums in Eqs.~\eqref{eq:coh_contr} and \eqref{eq:occ_contr}. Putting everything together, we obtain
	\begin{align}
	\overline{G}(\tau)&=\frac{|\mathcal{E}_0|^2}{4\hbar^2}\sum_{m,n,q,p} \rho_{p,p}^GM_n^m M_q^{m*}M_q^pM_n^{p*}e^{i\omega_{\rm ZPL}\tau}e^{-i(m-n)\omega\tau}\left[\frac{\gamma_{\rm pd}+\gamma_{\rm xd}}{2}+i\omega(\kappa-n+p)\right]^{-1}\notag\\
	&\qquad\times\left[\frac{\gamma_{\rm pd}+\gamma_{\rm xd}}{2}-i\omega(\kappa-q+p)\right]^{-1}\left[e^{i(\kappa+p-n)\omega \tau}+\frac{\gamma_{\rm pd}e^{-(\gamma_{\rm pd}+\gamma_{\rm xd})\tau/2}}{\gamma_{\rm xd}+i\omega(q-n)}\right]\,.
	\end{align}
	\section{Properties of the FC factors and relation to the phonon Wigner function}\label{app:FC}
	Here, we will give some analytical properties of the FC factors, which are defined by~\cite{maguire2019environmental}
	\begin{equation}
	M_m^n=\bra{m}B_+\ket{n}=\bra{m} D(\gamma)\ket{n}=M_m^n(\gamma)\,,
	\end{equation}
	which is the overlap of a Fock state displaced by $\gamma$ in phase space  $D(\gamma)\ket{n}$ with another undisplaced Fock state $\bra{m}$. One obvious property is
	\begin{equation}\label{eq:FC_cc}
	M_m^n(\gamma)^*=M_n^m(-\gamma)\,.
	\end{equation}
	They are also related to the symmetrically ordered characteristic function used to calculate Wigner functions of a single harmonic oscillator mode, e.g., in quantum optics~\cite{raimond2006exploring}, defined by
	\begin{equation}
	C_s^{\qty[\rho]}(\gamma)=\expval{D(\gamma)}=\Tr\qty[\rho D(\gamma)]\,.
	\end{equation} 
	The corresponding Wigner function is defined as the two-dimensional Fourier transform of the symmetrically ordered characteristic function
	\begin{align}\label{eq:def_wigner_1}
	W^{[\rho]}(\alpha)&=\frac{1}{\pi^2}\int \dd^2\lambda\, C_s^{[\rho]}(\lambda)\exp(\alpha\lambda^*-\alpha^*\lambda)=\sum_{m,n}\rho_{m,n}\frac{1}{\pi^2}\int \dd^2\lambda\, C_s^{[\ket{m}\bra{n}]}(\lambda)\exp(\alpha\lambda^*-\alpha^*\lambda)\notag\\
	&=\sum_{m,n}\rho_{m,n}\frac{1}{\pi^2}\int \dd^2\lambda\, M_n^m(\lambda)\exp(\alpha\lambda^*-\alpha^*\lambda)\equiv\sum_{m,n}\rho_{m,n}W^{[\ket{m}\bra{n}]}(\alpha)\,
	\end{align}
	with $\rho_{m,n}=\Tr(\bra{m}\rho\ket{n})$. Another representation of the Wigner function is given by
	\begin{align}
	W^{[\rho]}(\alpha)&=\frac{2}{\pi}\Tr\qty[D(-\alpha)\rho D(\alpha)\mathcal{P}]=\sum_{m,n}\rho_{m,n}\frac{2}{\pi}\Tr\qty[D(-\alpha)\ket{m}\bra{n}D(\alpha)\mathcal{P}]=\sum_{m,n}\rho_{m,n}W^{[\ket{m}\bra{n}]}(\alpha)\,.
	\end{align}
	Here, $\mathcal{P}$ is the phonon parity operator
	\begin{equation}
	\mathcal{P}=\exp(i\pi b^{\dagger}b)\,.
	\end{equation}
	Thus the Wigner function representation of the coherences $\ket{m}\bra{n}$ can be calculated as
	\begin{align}
	W^{[\ket{m}\bra{n}]}(\alpha)&=\frac{2}{\pi}\Tr\qty[\mathcal{P}D(-\alpha)\mathcal{P}^{\dagger}\mathcal{P}\ket{m}\bra{n}D(\alpha)]=(-1)^m\frac{2}{\pi}\Tr\qty[D(\alpha)\ket{m}\bra{n}D(\alpha)]\notag\\
	&=(-1)^m\frac{2}{\pi}\bra{n}D(2\alpha)\ket{m}=(-1)^m\frac{2}{\pi}M_n^m(2\alpha)=(-1)^{m-n}W^{[\ket{n}\bra{m}]}(-\alpha)^*\,.
	\end{align}
	In this way, we do not need to perform the two-dimensional Fourier transform in Eq.~\eqref{eq:def_wigner_1}, but can determine the Wigner function of our phonon mode directly from the FC factors. Now we continue by giving an analytical expression for the FC factors \cite{cahill1969ordered}
	\begin{align}
	M_n^m(\gamma)&=\bra{n}D(\gamma)\ket{m}=\exp(-\frac{|\gamma|^2}{2})\sum_{p,q}\frac{\gamma^q\qty(-\gamma^*)^p}{p!q!}\bra{n}\qty(b^{\dagger})^qb^p\ket{m}\notag\\
	&=\exp(-\frac{|\gamma|^2}{2})\sum_{p=0}^m\sum_{q=0}^n\frac{\gamma^q\qty(-\gamma^*)^p}{p!q!}\sqrt{\frac{m!n!}{(m-p)!(n-q)!}}\delta_{n-q,m-p}\,.
	\end{align}
	Due to Eq.~\eqref{eq:FC_cc} it is possible to restrict the calculation to the case $n\geq m$. The Kronecker delta in the expression for the FC factor then assures $q\geq p$\,. From now on, we assume $n\geq m$:
	\begin{align}\label{eq:M_ngeqm}
	M_n^m(\gamma)\Big|_{n\geq m}&=\exp(-\frac{|\gamma|^2}{2})\sum_{p=0}^m\sum_{q=0}^n\gamma^{q-p}\frac{\gamma^p\qty(-\gamma^*)^p}{p!q!}\sqrt{\frac{m!n!}{(m-p)!(n-q)!}}\delta_{n-q,m-p}\notag\\
	&=\exp(-\frac{|\gamma|^2}{2})\gamma^{n-m}\sqrt{\frac{m!}{n!}}\sum_{p=0}^m\sum_{q=0}^n(-1)^p\frac{|\gamma|^{2p}}{p!q!}\frac{n!}{(m-p)!}\delta_{n-q,m-p}\notag\\
	&=\exp(-\frac{|\gamma|^2}{2})\gamma^{n-m}\sqrt{\frac{m!}{n!}}\sum_{p=0}^m(-1)^p\frac{|\gamma|^{2p}}{p!}\frac{n!}{(n-m+p)!(m-p)!}\notag\\
	&=\exp(-\frac{|\gamma|^2}{2})\gamma^{n-m}\sqrt{\frac{m!}{n!}}\sum_{p=0}^m(-1)^p\frac{|\gamma|^{2p}}{p!}\binom{n}{m-p}\notag\\
	&=\exp(-\frac{|\gamma|^2}{2})\gamma^{n-m}\sqrt{\frac{m!}{n!}}\sum_{p=0}^m(-1)^p\frac{|\gamma|^{2p}}{p!}\binom{m+(n-m)}{m-p}\notag\\
	&=\exp(-\frac{|\gamma|^2}{2})\gamma^{n-m}\sqrt{\frac{m!}{n!}}L_{m}^{(n-m)}(|\gamma|^2)\,.
	\end{align}
	Here $L_m^{(n-m)}(x)$ is the generalized Laguerre polynomial~\cite{NIST:DLMF}. Using Eq.~\eqref{eq:FC_cc}, we get for the case of $m\geq n$
	\begin{align}\label{eq:M_mgeqn}
	M_n^m(\gamma)\Big|_{m\geq n}=M_m^n(-\gamma)^*\Big|_{m\geq n}=(-1)^{m-n}\exp(-\frac{|\gamma|^2}{2})\qty(\gamma^*)^{m-n}\sqrt{\frac{n!}{m!}}L_{n}^{(m-n)}(|\gamma|^2)\,.
	\end{align}
	For the absolute value of the FC factors, we actually get a simpler symmetry property than Eq.~\eqref{eq:FC_cc}, as can be seen from Eq.~\eqref{eq:M_ngeqm} and Eq.~\eqref{eq:M_mgeqn}, reading
	\begin{equation}\label{eq:abs_FC_symmetry}
	\qty|M_n^m(\gamma)|=\qty|M_m^n(-\gamma)|=\qty|M_m^n(\gamma)|\,.
	\end{equation}
	In the regime of weak phonon coupling, i.e., $\gamma\ll 1$, keeping only the lowest order in $|\gamma|$, we obtain
	\begin{align}\label{eq:FC_weak}
	\qty|M_n^m(\gamma)|&\approx \frac{|\gamma|^{|m-n|}}{|m-n|!}\sqrt{\frac{\max(m,n)!}{\min(m,n)!}}\,.
	\end{align}
	\section{Asymptotic behavior of the semiclassical approximation}\label{app:semiclassical}
	To evaluate the asymptotic behavior of the semiclassical approximation for large coherent state amplitudes $\alpha$, we start by considering the series from Eq.~(35) in the main text, which reads
	\begin{align}
	S&(|\alpha|^2)=\sum_i\frac{\max(i+\kappa,i+\kappa-k)!}{\min(i+\kappa,i+\kappa-k)!}\frac{\max(i+\kappa,i)!}{\min(i+\kappa,i)!}\frac{|\alpha|^{2i}}{i!}\,.
	\end{align}
	For the different cases for $k$ and $\kappa$, this can be written in terms of the generalized hypergeometric functions $_pF_q$ as~\cite{NIST:DLMF}
	\begin{equation}
	S(|\alpha|^2)=\begin{cases}
	\frac{\kappa!\kappa!}{(\kappa-k)!}\phantom{}_2F_2(\kappa+1,\kappa+1;1,\kappa-k+1;|\alpha|^2)&k\geq 0\,,\ \kappa\geq 0\\
	|\alpha|^{2(|k|+|\kappa|)}e^{|\alpha|^2}&k\geq 0\,,\ \kappa<0\\
	(\kappa-k)!_1F_1(\kappa-k+1;1;|\alpha|^2)&k<0\,,\ \kappa\geq 0\\
	|\alpha|^{2|\kappa|}|k|!_1F_1(|k|+1;1;|\alpha|^2)&k<0\,,\ \kappa<0\,.
	\end{cases}
	\end{equation}
	In the limit $|\alpha|^2\rightarrow\infty$, all four cases behave as~\cite{NIST:DLMF}
	\begin{equation}
	\lim\limits_{|\alpha|^2\rightarrow\infty} S(|\alpha|^2)\sim |\alpha|^{2(|k|+|\kappa|)}e^{|\alpha|^2}\,.
	\end{equation}
	
	\medskip
	
	%